\documentclass[prd,showpacs,nofootinbib,eqsecnum,amsfonts,amssymb,amsmath,preprintnumbers]{revtex4}
\usepackage[pdftex]{graphicx,color}
\usepackage{dsfont}
\usepackage{bm}
\usepackage{textcomp}

\newcommand{\ma}[1]{{\mathrm{#1}}}

\newcommand{\calH}{{\cal H}}
\newcommand{\calL}{{\cal L}}
\newcommand{\calM}{{\cal M}}
\newcommand{\calD}{{\cal D}}

\newcommand{\calP}{{\cal P}}
\newcommand{\calC}{{\cal C}}

\newcommand{\calS}{{\cal S}}
\newcommand{\calQ}{{\cal Q}}
\newcommand{\calT}{{\cal T}}
\newcommand{\pa}{{\partial}}

\newcommand{{\Pone}}[1]{{#1}^{\prime}{}}
\newcommand{{\Ptwo}}[1]{{#1}^{\prime \prime}{}}
\newcommand{\dd}{{\rm d}}
\newcommand{\ii}{{\rm i}}

\newcommand{\BBdis}{{I_{\rm BB}}}
\newcommand{\BBdisB}{{I_{\rm BB} \left( \frac{q}{a \bar T} \right)}}
\newcommand{\parq}{{\frac{\partial}{\partial \ln q}}}
\newcommand{\BBparq}{{\frac{\pa I_{\rm BB}}{\pa \ln q}}}
\newcommand{\parqtwo}{{\frac{\partial^2}{\partial \ln q^2}}}
\newcommand{\parqof}[1]{{\frac{\partial #1}{\partial \ln q}}}

\newcommand{\bepsilon}{{\boldsymbol{\epsilon}}}
\newcommand{\bxi}{{\boldsymbol{\xi}}}
\newcommand{\bzeta}{{\boldsymbol{\zeta}}}
\newcommand{\RR}{\mathbb{R}}

\begin{document}

\title{Second order Boltzmann equation : gauge dependence and gauge invariance}

\preprint{YITP-13-23}

\author{Atsushi Naruko$^{1}$, Cyril Pitrou$^{2,3}$, Kazuya Koyama$^{4}$,
 Misao Sasaki$^{5}$}

\affiliation{$1$ APC (CNRS-Universit\'e Paris 7),
 10 rue Alice Domon et L\'eonie Duquet, 75205 Paris Cedex 13, France
}

\affiliation{$2$ Institut d'Astrophysique de Paris,
Universit\'e Pierre~\&~Marie Curie - Paris VI,
CNRS-UMR 7095, 98 bis, Bd Arago, 75014 Paris, France
}

\affiliation{$3$ Sorbonne Universit\'es, Institut Lagrange de Paris,
  98 bis bd Arago, 75014 Paris, France
}

\affiliation{$4$ Institute of Cosmology and Gravitation,
 University of Portsmouth, Dennis Sciama Building Burnaby Road,
 Portsmouth PO1 3FX, UK
}

\affiliation{$5$ Yukawa Institute for Theoretical Physics, Kyoto University,
 Kyoto 606-8502, Japan
}

\date{May 8, 2013} 


\begin{abstract}
In the context of cosmological perturbation theory, we derive the
second order Boltzmann equation describing the evolution of the
distribution function of radiation without a specific gauge choice.
The essential steps in deriving the Boltzmann equation are revisited and extended
given this more general framework: i)
the polarisation of light is incorporated in this formalism by
using a tensor-valued distribution function; ii) the importance of a choice of the tetrad field to define the local
inertial frame in the description of the distribution function is
emphasized; iii) we perform a separation between temperature and
spectral distortion, both for the intensity and for
polarisation for the first time; iv) the gauge dependence of all perturbed quantities
that enter the Boltzmann equation is derived, and this enables us to check
the correctness of the perturbed Boltzmann equation by explicitly
showing its gauge-invariance for both intensity and polarization. We
finally discuss several implications of the gauge dependence for the observed temperature.
\end{abstract}

\maketitle

\section{Introduction}

The non-Gaussianity in the Cosmic Microwave Background (CMB) has been one of the hottest topics in cosmology because it could open a new window for probing the primordial universe.
Recent CMB observations, especially WMAP~\cite{Komatsu:2010fb} and
 Planck~\cite{Ade:2013ydc}, have confirmed to a very high accuracy that the primordial curvature
perturbations have a nearly scale invariant initial power spectrum and
the associated statistics is nearly Gaussian. These observations
are consistent with the predictions of an early inflationary era
driven by a single slow-rolling scalar field.

The possibility of non-Gaussianity in the primordial curvature
perturbations was discussed for the first time quantitatively by Komatsu and Spergel~\cite{Komatsu:2001rj}. They parameterized the level of non-Gaussianity in the potential $\Phi(x)$ (the curvature potential in the Newton or Poisson gauge)  by
 \begin{equation}
 \Phi (x) = \Phi_\ma{L} (x) + f_\ma{NL}^\ma{local} \Bigl[ \Phi^2_\ma{L} (x)
 - \langle \Phi_\ma{L}^2 (x) \rangle \Bigr] \,.
 \end{equation}
Here, $\Phi_\ma{L} (x)$ denotes the Gaussian part of the perturbation,
or in perturbation theory its linear part, and $\langle \cdots \rangle$
designates the statistical average. This type of non-Gaussianity
leads to a non-vanishing three point correlation function, or
equivalently in reciprocal space to a non-vanishing bispectrum.
The prediction for the possible values of this parameter
 $f^\ma{local}_{\ma{NL}}$ from a phase of single-field slow-roll inflation
was first performed by Maldacena~\cite{Maldacena:2002vr}, and it was shown that it is of order of the slow-roll parameters and thus highly
suppressed. Hence, if a $f^\ma{local}_{\ma{NL}}$ of order unity or greater is detected, this simplest model of inflation will be ruled out.

Unfortunately, the interpretation of the measured non-Gaussianity is not so
straightforward because we do not observe directly the primordial non-Gaussianity in
the curvature perturbation but its effect on the CMB
fluctuations. Therefore, to relate the primordial curvature perturbation to CMB,
we first have to compute the evolution of perturbations after
inflation. These effects can be split unambiguously in two parts: i) a
linear transfer that cannot create a non-Gaussian signal if the
initial conditions are purely Gaussian, and ii) a non-linear transfer
that generates a non-Gaussian signal in the observables even if the
initial conditions are purely Gaussian. The resulting non-Gaussian signal from i)
is often called {\it primordial non-Gaussianity} and all the possible sources
of non-linear evolutions which enter the category ii) are called
{\it secondary non-Gaussianity}.
Recently the Planck collaboration provided a constraint $f^\ma{local}_{\ma{NL}} = 2.7 \pm 5.8$ \cite{Ade:2013ydc}.
This result was obtained by subtracting one of the secondary non-Gaussianities that
arises from the correlation between the lensing and integrated Sachs-Wolfe effect. This clearly demonstrates the importance of subtracting all the secondary non-Gaussianity consistently in order to obtain an accurate constraint on the primordial non-Gaussianity.

The evolution of the perturbations on super-horizon scales is well understood,
even fully non-linearly using either a covariant approach~\cite{Langlois:2005ii,Langlois:2005qp,Enqvist:2006fs,Pitrou:2007xy},
or a separate universe approach with the so-called $\delta N$
formalism~\cite{Comer:1994np,Kolb:2004jg,Lyth:2004gb}, since it leads to a conservation law for the curvature perturbations in the case of adiabatic perturbations.
On the other hand, the evolution for modes below the horizon scale is
not so simple analytically, especially at the non-linear order, and
the use of a kinetic description cannot be avoided on small scales
since radiation starts to develop an anisotropic stress.
In order to obtain numerical results for the non-linear evolution,
we need to derive and solve without approximations the coupled system of non-linear a) Einstein equation for the metric, b) conservation and Euler equations for fluids and c)
Boltzmann equation for radiation (photons and neutrinos).
Note that the conservation and Euler equations can always be deduced from the
lowest moments of the Boltzmann equation, and the full set of equations
is often only referred to as {\it Einstein-Boltzmann system} of equations.

In order to follow this roadmap, the second order Boltzmann equation
was written down in the Poisson gauge in
Refs.~\cite{Bartolo:2006cu,Bartolo:2006fj,Pitrou:2008hy,Pitrou:2008ut,Pitrou:2010sn,Beneke:2010eg}.
The gauge dependence of the distribution function was obtained at linear
order~\cite{Durrer:1993db} and then at second order~\cite{Pitrou:2007jy} but leaving aside the problem of polarisation. It was then extended to include polarised light in Ref.~\cite{Pitrou:2008hy}.
The system of equations was then solved numerically in Fourier space in Poisson gauge in~\cite{Pitrou:2010sn}, and it was reported that the secondary effects around the last-scattering surface could mimic a primordial signal of $f_\ma{N L}^\ma{local} \sim 4$.
Recently there have been a huge progress in improving the numerical
calculations and clarifying the amplitude of various secondary
non-Gaussianities at recombination~\cite{Huang:2012ub,Su:2012gt,Pettinari:2013he}, and a consensus
emerged that when including all the non-linear effects around
recombination \emph{and} the integrated early effects after
recombination, it could mimic a primordial signal of $f_\ma{N L}^\ma{local} \sim 0.8$, as
expected from analytic approximations~\cite{Creminelli:2011sq}.

The description of the spectral dependence of the distribution function
is also crucial at second order.
Indeed, at first order there are no spectral distortions and
the perturbation of the photon distribution function
can be understood as a single, spectrum-independent temperature
fluctuation. However, at second order, there appears a deviation from the
Planck distribution, resulting in a continuum of spectral
distortions, which in principle must superimpose the thermal Sunyaev-Zeldovich effect~\cite{Ade:2013qta}. To describe this distortion,
 we use a direction and position dependent Compton $y$
parameter~\cite{Stebbins:2007ve,Pitrou:2009bc}, and also introduce a
similar tensor-valued variable to describe the distortion in the
polarisation, thus extending the formalism introduced
 in Ref.~\cite{Pitrou:2009bc}.
In this paper, we derive the second order Boltzmann equation without
restricting to a specific gauge, and including
polarisation. Reflecting on the above, our motivation is two-fold.

First, since the structure of the second order Einstein and
Boltzmann equations depends very much on a choice of the gauge,
we have to find a gauge in which we can numerically solve this system
accurately and quickly.
Therefore, it is preferable not to specify the gauge from the beginning
but to formulate the equations without specifying it. We can impose
different gauge restrictions in their final form to explore the
stability and efficiency of the numerical integration.
Second, we would like to check the equations derived in
Refs.~\cite{Pitrou:2008hy,Pitrou:2010sn,Beneke:2010eg}. As a direct
check, we recover them in the specific case of the Poisson
gauge. Then, as an indirect check, we revisit the transformation properties
of the distribution function and the metric perturbations and confirm
that the perturbed Boltzmann equation is gauge-invariant up to second
order in perturbations, thus increasing our confidence in the rather
lengthy derivation.

In Ref.~\cite{Pettinari:2013he}, it was found that the inclusion or omission of certain line of sight terms can make a large impact on the estimation of the bias to the primordial non-Gaussianity due to the secondary non-Gaussianity. In Refs~\cite{Huang:2012ub,Pettinari:2013he} all physical effects were included except for lensing and time-delay. These time-integrated effects require a separate analysis because at later times small-scale multipoles get excited and numerically it is very difficult to evolve the equations. In this paper, we point out that the separation of these effects  depends on a gauge. Given that the lensing-ISW cross correlation gives the largest bias to the primordial local type non-Gaussianity, one should bear this gauge dependence in mind when separating these time integrated effects in the calculations.

The choice of the gauge and the associated choice of the tetrad field for
the distribution function is also crucial in the interpretation of the quantities
as observables. These subtle details do not affect our
interpretation of observables in the linear theory since it is only relevant for
the monopole and the dipole. However it is no longer the case at second order
in perturbations. We must understand the transformation properties of the
distribution function under a gauge transformation or a change of the inertial frame
and determine what is a gauge and a choice of inertial frame that is related
to CMB experiments.

The structure of this paper is as follows. In section~\ref{sec:def}, we give the definitions of the variables that we use for the metric, momentum and distribution function.
Especially, to express the perturbation of the metric, we use a geometrical
$(3+1)$ decomposition, or the ADM~\cite{Arnowitt:1962hi} parametrisation of the metric.
At first order, there is no particular advantage in using this formalism, but various expressions are simplified at second order for the choice of the inertial frame that we make. In section~\ref{sec:Boltz}, we derive the second order Boltzmann equation
with polarisation without restricting to a specific gauge.
In section~\ref{sec:gauge}, we discuss the gauge dependence of the variables.
We carefully investigate the gauge transformation of the metric,
momentum and the distribution function.
We then check explicitly the gauge invariance
of the perturbed Boltzmann equation up to second order as a
consistency test.
Finally, in Section \ref{sec:conc}, we summarize our
results and we comment briefly on the relevance of our formalism for
the observed CMB anisotropy. Useful technical details are gathered in the appendices.

\section{Definitions}
\label{sec:def}

In this section we build all the tools which are used for the
description of polarized radiation in cosmology. We first review
briefly the parametrization of cosmological perturbations, and explain
how a photon momentum can be uniquely described by its energy and
direction once a suitable tetrad choice has been made. We then
introduce the tensor-valued distribution function which is used to
treat statistically a gas of polarized photons, and which is the key
object in the Boltzmann equation, and we finally present how it can
be decomposed into its main spectral components.

\subsection{Spacetime coordinates and local inertial frame}
\label{ssec:deftet}

We shall use the ADM formalism to write down the expression of the perturbed metric where the metric can be decomposed as
\begin{align}\label{MetricADM}
 \dd s^2 &= a^2(\eta) \Bigl[ - N^2 \dd \eta^2
 + \gamma_{ij} (\dd x^i + \beta^i \dd \eta)
 (\dd x^j + \beta^j \dd \eta) \Bigr] \nonumber\\
 &= a^2(\eta) \Bigl[ - (N^2 - \gamma_{i j} \beta^i \beta^j) \dd \eta^2
 + 2 \gamma_{ij} \beta^j \dd x^i \dd \eta
 + \gamma_{ij} \dd x^i \dd x^j \Bigr] \,,
\end{align}
 where $N$ is the lapse function, $\beta^i$ is the shift vector,
 $\gamma_{i j}$ is the spatial metric, and indices of the Latin
  type
 ($i,j,k\cdots$) run from $1$ to $3$.
To describe the perturbations around
 the flat Friedmann-Lema\^itre-Robertson-Walker (FLRW) space-time,
 perturbation variables, $\alpha$ and $h_{i j}$, are introduced as
\begin{equation}\label{DefPertNNi}
 N \equiv 1 + \alpha \,, \qquad
 \gamma_{i j} \equiv \delta_{i j} + 2 h_{i j} \,.
\end{equation}
For simplicity, we use the following definitions
\begin{equation}
\beta_i\equiv \delta_{ij} \beta^j \,, \quad
 h^i{}_j \equiv \delta^{ik} h_{kj} \,, \quad
 h^{ij} \equiv \delta^{ik} \delta^{jl}h_{kl} \,,
\end{equation}
where the spatial indices are raised and lowered
with $\delta_{ij}$ and $\delta^{ij}$, rather than with $\gamma_{ij}$ and $\gamma^{ij}$.
As is clearly seen below, the ADM form of the metric perturbation will simplify
the expressions of the perturbed Boltzmann equation.
Any perturbation $X$ will be expanded into its first and second order parts as
\begin{equation}
 X= X^{(1)} + \frac{1}{2} X^{(2)}\,.
\end{equation}
The relations between the ADM variables and the usual definitions of cosmological perturbations are provided in Appendix~\ref{app:ADM}.

The Boltzmann equation is better formulated by explicitly using a
local inertial frame at every point of the space-time and this can be achieved by using a tetrad field. It is a set of four vector fields which satisfy
\begin{equation}
 \eta_{(a) (b)} = g_{\mu \nu} e_{(a)}{}^\mu e_{(b)}{}^\nu \,, \qquad
 g_{\mu \nu} = \eta_{(a) (b)} e^{(a)}{}_\mu e^{(b)}{}_\nu \,.
\end{equation}
These conditions determine the choice of tetrad only up to rotations and boosts.
Here the following particular tetrads are chosen up to second order accuracy
\begin{align}
 e^{(0)}{}_\mu &= a (- N, 0, 0, 0) \,,\qquad
 e^{(i)}{}_\mu = a \left( \beta^i + h^i{}_j \beta^j,
 \delta^i{}_j + {h^i}_j - \frac{1}{2} h^{i k} h_{k j} \right) \,,
\label{tetrads}
\end{align}
and the inverse tetrads are given by
\begin{align}
 e_{(0)}{}^\mu = - e^{(0) \mu} &= - \frac{1}{a} \left( \frac{1}{N}, - \frac{\beta^i}{N} \right) \,, \qquad
 e_{(i)}{}^\mu = \frac{1}{a} \left[ 0, \delta_{(i) (k)} \left( \delta^{j k}
 - h^{j k} + \frac{3}{2} h^{j l} h_l{}^k \right) \right] \,.
\end{align}
The time-like tetrad is chosen to be orthogonal to the constant time
hypersurfaces since ${\bm e}^{(0)} \propto \dd \eta$.
As for the spatial tetrads, this choice corresponds to asking that there is no rotation between the background and the perturbed tetrads~\cite{Pitrou:2007jy}.

\subsection{Momentum}

To facilitate the separation between the magnitude of the momentum and its direction in a covariant manner, let us consider the projection of the momentum of photon $p^\mu$ onto the set of tetrads,
 \begin{align}
 p^{(a)} = e^{(a)}{}_\mu p^\mu \,.
 \end{align}
We introduce the conformal momentum of photon rather than the physical momentum $p^{(a)}$
\begin{equation} \label{defpP}
 q^{(a)} \equiv a p^{(a)} \,.
\end{equation}
Since the momentum of photon satisfies the null condition $p^\mu p_\mu = 0$,
or equivalently $q^{(a)} q_{(a)} = 0$, only three components among
four are independent, that is
 \begin{align}
 q^{(a)} q_{(a)} = 0 \,, \quad \Leftrightarrow \quad
 (q^{(0)})^2 = \delta_{(i) (j)} q^{(i)} q^{(j)}  \,.
 \end{align}
Thus the three spatial components $q^{(i)}$ can be regarded as such independent variables.
Furthermore, $q^{(i)}$ can be decomposed into its magnitude $q$ and direction $n^{(i)}$ as
 \begin{align}
 q \equiv \sqrt{\delta_{(i) (j)} q^{(i)} q^{(j)}} = |q^{(0)}| \,, \qquad
 n^{(i)} \equiv \frac{q^{(i)}}{q} \,.
 \end{align}
Physically the above $q$ can be understood as the conformal (re-scaled) energy, $q = a E_\ma{phys}$, seen by an observer orthogonal to time constant hypersurfaces.

From Eq.~(\ref{defpP}), the components of momentum $p^{\mu}$ are expressed as functions of $(q, n^{(i)})$ up to the second order as
\begin{subequations}
\begin{align}\label{p0pi}
 p^0 &= \frac{q}{a^2} (1 - \alpha + \alpha^2) \,, \\
 p^i &= \frac{q}{a^2} \left( n^{(i)} - \beta^i - {h^i}_j n^{(j)}
 + \alpha \beta^i + \frac{3}{2} h^{i k} h_{k j} n^{(j)} \right) \,.
\label{momentum}
\end{align}
\end{subequations}
Conversely, ($q, n^{(i)}$) are given by the components of momentum as
\begin{subequations}
\begin{align}
 q &= a^2 (1 + \alpha) p^0 \,,
\label{def:q} \\
 n^{(i)} &= \left[ (1 - \alpha + \alpha^2) \delta^i{}_j + (1 - \alpha) h^i{}_j
 - \frac{1}{2} h^{ik} h_{k j} \right] \frac{p^j}{p^0}
 + (1 - \alpha) \beta^i + \beta^j {h^i}_j \,.
\label{def:n^i}
\end{align}
\end{subequations}

One can introduce a projection operator in terms of
 $e^{(0)}{}_\mu$ and $n_\mu$.
The projection operator, often called the screen projector, is defined as
\begin{align}
 S_{\mu \nu}
 &\equiv g_{\mu \nu} + e^{(0)}{}_\mu e^{(0)}{}_\nu - n_\mu n_\nu \,,
\end{align}
 where $n^\mu$, the direction vector of photon, is defined by
 \begin{align}
 n^\mu \equiv e_{(i)}{}^\mu n^{(i)} \,.
 \end{align}
Clearly $S_{\mu \nu}$ is a projection of the tangent space onto a two dimensional plane
orthogonal to both $e^{(0)}{}_\mu$ and $n_\mu$ since $S^{\mu \nu} e^{(0)}{}_\mu$ and $S^{\mu \nu} n_\mu$ vanish. 
Its expression in tetrad components reduces necessarily to the
identity of the two-dimensional subspace which is left invariant by
the projector, that is
\begin{align}
 S_{(i) (j)} = \delta_{(i) (j)} - n_{(i)} n_{(j)} \,,\qquad
 S_{(0) (0)}=S_{(0) (i)}=0 \,.
\end{align}

\subsection{Distribution function for photons and Stokes parameters}
\label{ssec:f}

In order to describe the polarisation of radiation, we introduce a tensor-valued distribution function $f_{\mu \nu}$, which is complex valued and Hermitian.
The construction of this distribution function is discussed in Appendix~\ref{app:Dist}.
It is independent of the choice of the electromagnetic gauge and contains
 only four physical degrees of freedom since it satisfies the conditions
\begin{equation}
 f_{\mu\nu} e_{(0)}{}^\mu = f_{\mu\nu} e_{(0)}{}^\nu
 = f_{\mu\nu} n^\mu = f_{\mu\nu} n^\nu = 0 \,.
\end{equation}
Note that the distribution function depends on the observer's velocity,
$u^\mu \equiv e_{(0)}{}^{\mu}$, used in its definition. As long as no confusion
arises from such dependence, we omit to specify it.
In the case where this is needed, mainly when studying the transformation properties
of such a quantity, we shall use the notation $f_{\mu\nu}^{\bm{e}_{(0)}}$ to stress that
the tensor-valued distribution function is dependent on the observer's velocity and thus on the choice of the tetrad field.

The four degrees of freedom can be extracted by decomposing $f_{\mu \nu}$
into a trace part, a symmetric traceless part and an antisymmetric part as
\begin{equation}
 f_{\mu \nu}  \equiv \frac{1}{2} I
 S_{\mu \nu} + P_{\mu \nu}+ \frac{\ii}{2} \epsilon_{\rho \mu \nu \sigma}
 e_{(0)}{}^\rho n^\sigma V \,,
 \label{Pol}
\end{equation}
where the antisymmetric tensor is defined by
\begin{gather}
 \epsilon_{\alpha \beta \gamma \delta} = \epsilon_{[ \alpha \beta \gamma \delta] } \,,
 \quad \epsilon_{0 1 2 3} = \sqrt{- g} \,, \qquad {\rm or} \qquad
 \epsilon_{(0)(1)(2)(3)}=-\epsilon^{(0)(1)(2)(3)} = 1 \,.
\end{gather}
{\it I} is the intensity and $V$ is the degree of circular polarisation.
$P_{\mu \nu}$ encodes the two degrees of linear polarisation
(so called $Q$ and $U$ Stokes parameters). All these functions, together with
the original tensor-valued distribution function, are functions of the position
on space-time $x^\mu=(\eta,x^i)$ and on the point in tangent space.
This point in the tangent space can be chosen to be parametrized either by the
components $p^\mu$ in the basis canonically associated with the coordinates system, or alternatively by their Cartesian counterparts $p^{(a)}$.
In fact we will choose to parametrize the tangent space by the
components of the conformal momentum in tetrad space, $q^{(i)} = a
p^{(i)}$, expressed in their spherical coordinates $q$ and $n^{(i)}$,
as this leads to the most simple form for the Boltzmann equation as we
shall see further.

\subsection{Spectral distortion}

On the background space-time, the distribution function, which is characterized only by the intensity $I$, is given by a Planck distribution whose temperature $\bar T$ depends only on $\eta$ due to the symmetries of the FLRW universe. As we will check later, the background temperature scales as $\propto 1/a$. We thus have
\begin{equation}
\bar I(\eta,q) = \BBdis\left[\frac{q}{a(\eta)\bar T(\eta)}\right] \,, \quad
 \text{with} \quad \BBdis (x) \equiv \frac{2}{( e^x - 1 )} \,.
\end{equation}

At first order in perturbation, the fluctuation of intensity can be described
as a fluctuation of temperature $\delta T$ which is independent of $q$.
There are two reasons for this. First, as we shall discuss further, gravitational interactions do not induce spectral distortions in the sense that they shift all wavelengths by the same ratio. Second, the collisions at linear order in perturbation do not induce spectral distortions and the redistribution of the photon directions resulting
from it can be described by a direction dependent temperature. A similar procedure can be followed for the description of polarisation at first order.

However at second order the situation becomes more complicated since the Compton scattering at this order of perturbation induces spectral distortions which cannot be reabsorbed
in a simple direction dependent temperature. As a result, the photon distribution is not described by a Planck distribution function, but fortunately it is sufficient to use two
direction dependent quantities. The first remains the temperature and the second describes
the type of spectral distortion generated at second order. Actually in general, at the $n$-th order, $n$ directional dependent functions would be needed~\cite{Stebbins:2007ve,Pitrou:2009bc} to characterize fully the spectrum.

In order to parametrize this distortion, we introduce on top of the
temperature $T$, the so-called Compton $y$ parameter. In this section
we will omit the dependence of all quantities on the coordinates
$x^\mu$ and we will focus on the dependence on the tangent space
coordinates $(q,n^{(i)})$. The distribution function can be expanded around a Planck distribution in the so-called Fokker-Planck expansion
as ~\cite{Stebbins:2007ve}
\begin{align} \label{defy}
 I \Bigl( q,n^{(i)} \Bigr)
 &\simeq \BBdis \left( \frac{q}{a T} \right)
 + y \bigl( n^{(i)} \bigr) q^{-3} \parq \left[
 q^3 \parq \BBdis \left( \frac{q}{a T} \right) \right] \notag\\
 &= \BBdis \left( \frac{q}{a T} \right) + y \bigl( n^{(i)} \bigr)
 \calD_q^2\BBdis \left( \frac{q}{aT} \right) \,,
\end{align}
 where
\begin{equation}
 \calD_q^2 \equiv q^{-3} \parq \left( q^3 \parq \right)
 = \parqtwo+3 \parq \,.
\end{equation}
Because the number density of photon is given by $n \propto a^{-3}\int I q^2 \dd q$, the $y$ term does not contribute to the photon number density and the temperature $T$ is
the temperature of the black-body that would have the same number density (see Ref.~\cite{Pitrou:2010sn} for a discussion on other possible definitions for the temperature) and we call it here {\it number density temperature}.
It can be expanded around the background temperature as
\begin{equation}
 T \bigl( n^{(i)} \bigr)
 \equiv \bar T(\eta) \left[ 1 + \Theta \bigl( n^{(i)} \bigr) \right] \,.
\end{equation}
Note that the expansion~(\ref{defy}) is not the same as Eq. ($11$) nor
 Eq. ($15$) of Ref.~\cite{Stebbins:2007ve}.
Indeed, the temperature of the Planck spectrum around which we expand is
neither the physically motivated logarithmic averaged temperature of Ref.~\cite{Stebbins:2007ve} nor a fiducial temperature, but another physically motivated temperature (the number density temperature) that suits better to describe the spectral distortion of the type that appears in CMB.

However, when performing perturbations in cosmology, we need to refer to the background space-time temperature $\bar T$, not to the local number density temperature. Thus it
is convenient to expand the distribution function around a Planck
distribution at $\bar T$ rather than $T$. Expanding Eq.~(\ref{defy})
in $\Theta$ up to the second order, we obtain the expansion as
\begin{align}
 I &= \BBdisB - \bigl( \Theta+\Theta^2 \bigr) \parq \BBdisB
 +\left( y + \frac{1}{2} \Theta^2 \right) \calD_q^2 \BBdisB \,,
\label{intensity}
\end{align}
where we used the fact that $y$ is at least a second order quantity.
Here, in order to simplify the notation, it is implied that $\Theta$
and $y$ depend on $x^\mu$ and $n^{(i)}$. For a given $I$, the spectral
components $\Theta$ and $y$ can be extracted by performing different
types of integrals on $q$ (see appendix~\ref{AppExtraction} for details).
This expansion is similar to Eq. ($11$) of Ref.~\cite{Stebbins:2007ve}
when only second derivatives of the Planck distribution are kept.

Now we want to obtain a similar decomposition for polarisation.
Indeed, when dealing with polarisation we also need to expand its spectral
dependence in a way similar to what has been performed for the
intensity in Eqs.~(\ref{defy}) and (\ref{intensity}), that is we want
to separate the polarisation tensor into a spectral distortion
$Y_{\mu\nu}$ and non-distorted component $\calP_{\mu \nu}$. However,
this separation is slightly different given that there is no
polarisation on the background and hence there is no term corresponding to
the first term in Eq.~(\ref{intensity}). In the appendix of Ref.~\cite{Stebbins:2007ve}, it has been shown that the expansion should be
\begin{equation}\label{defyPbase}
 P_{\mu \nu} \Bigl( q, n^{(i)} \Bigr)
 \simeq - \calP_{\mu\nu} \bigl( n^{(i)} \bigr) \parq \BBdis
 \left( \frac{q}{a T} \right)
 + Y_{\mu \nu} \bigl( n^{(i)} \bigr) \calD_q^2 \BBdis
 \left( \frac{q}{a T} \right) \,,
\end{equation}
which is just a consequence of the fact that there is no background polarisation.
We will check that $Y_{\mu\nu}$ vanishes at first order as it is not generated by collisions at this order. Similarly to the expansion of the intensity part, we want to expand the distribution function around a Planck spectrum at the background temperature
$\bar T$ rather than the local number density temperature $T$. Thus we expand Eq.~(\ref{defyPbase}) in $\Theta$ up to first order to get
\begin{align}\label{defyP}
 P_{\mu \nu} &= - (1 + 3 \Theta) \calP_{\mu\nu} \parq \BBdisB
 + (Y_{\mu \nu} + \Theta \calP_{\mu\nu}) \calD_q^2 \BBdisB \,.
\end{align}
Again here, in order to simplify the notation, it is implied that
 $\Theta$, $\calP_{\mu \nu}$ and $Y_{\mu\nu}$ depend on $x^\mu$ and $n^{(i)}$.

\section{Boltzmann equation}
\label{sec:Boltz}

Now that we have all the tools at hand, we are ready to formulate
the Boltzmann equation  for polarized radiation in the cosmological context, and extract its
spectral components. This section is entirely dedicated to this
task. Given that the complete and detailed derivation can be rather
lengthy, all details which are not necessary in a first reading are
gathered in  Appendix~\ref{app:Boltz_pol}.
We first present the general expression of the Boltzmann equation for
a tensor-valued distribution function. Since the Boltzmann equation is
the description of how this distribution function evolves along a
photon geodesic, it is necessary to perturb the geodesic
equation up to second order. We then show how the Boltzmann equation can be split into its main spectral
components, that is into a temperature and a distortion. Finally we
write the explicit forms of the free-streaming part and the collision
part of the Boltzmann equation.

\subsection{Boltzmann equation}

The evolution of the tensor-valued distribution function is dictated
 by the Boltzmann equation~\cite{Tsagas:2007yx}
\begin{equation}\label{Evolfmunu}
 S_\mu{}^\rho S_\nu{}^\sigma \frac{\calD f_{\rho \sigma}}{\calD \lambda}
 = C_{\mu \nu} \,,
\end{equation}
where $\calD / \calD \lambda$ is the covariant derivative
 along a photon trajectory $x^{\mu}(\lambda)$ and the momentum and
 $C_{\mu \nu}$ is the associated collision term.
The explicit form of $\calD f_{\mu \nu}/ \calD \lambda$ is
\begin{equation}
 \frac{\calD f_{\mu \nu}}{\calD \lambda}
 \equiv \nabla_\rho f_{\mu \nu} \frac{\dd x^\rho}{\dd \lambda}
 + \frac{\pa f_{\mu \nu}}{\pa q^{(i)}} \frac{\dd q^{(i)}}{\dd \lambda} \,,
\end{equation}
 where $\nabla_\mu$ indicates a covariant derivative
 associated with $g_{\mu \nu}$.
Using spherical coordinates $(q,n^{(i)})$ instead of $q^{(i)}$
 for the momentum space, the Liouville operator, that is the l.h.s of
 Eq.~(\ref{Evolfmunu}), reads
\begin{equation}\label{DDLoperatortensor}
 S_\mu{}^\rho S_\nu{}^\sigma \frac{\calD f_{\rho \sigma}}{\calD \lambda}
 = S_\mu{}^\rho S_\nu{}^\sigma \nabla_\tau f_{\rho \sigma}
 \frac{\dd x^\tau}{\dd \lambda}
 + \frac{\pa f_{\mu \nu}}{\pa \ln q} \frac{\dd \ln q}{\dd \lambda}
 + D_{(i)} f_{\mu \nu} \frac{\dd n^{(i)}}{\dd \lambda} \,,
\end{equation}
 where $D_{(i)}$ is a covariant derivative for momentum.
The detail of the construction of such derivative is discussed
 in Appendix \ref{sapp:covp}.
Thanks to the operation of the projection $S_\mu{}^\rho S_\nu{}^\sigma$ onto
 $ \calD f_{\rho \sigma} / \calD \lambda$,
the left hand side of the Boltzmann equation can be decomposed into
the $I$, $V$ and $P_{\mu \nu}$ parts similarly to Eq.~(\ref{Pol});
\begin{gather}
 S_\mu{}^\rho S_\nu{}^\sigma \frac{\calD f_{\rho \sigma}}{\calD \lambda}
 = \frac{1}{2} L[I] S_{\mu \nu} + L[{\bf P}\,]_{\mu \nu}
 + \frac{\ii}{2} L[V] \epsilon_{\rho \mu \nu \sigma} e_{(0)}{}^\rho n^\sigma \,,
\end{gather}
where the corresponding Liouville operators are defined as
\begin{equation}\label{defofLs}
 L[I] \equiv \frac{\calD I}{\calD \lambda} \,, \qquad
 L[{\bf P} \, ]_{\mu \nu} \equiv  S_\mu{}^\rho S_\nu{}^\sigma
 \frac{\calD P_{\rho \sigma}}{\calD \lambda} \,, \qquad
 L[V]\equiv \frac{\calD V}{\calD \lambda} \,.
\end{equation}
Here, the operator $\calD / \calD \lambda$ on a scalar distribution
 function $f$, takes the simpler form
 \begin{equation}
 \frac{\calD f}{\calD \lambda} \Bigl( x^\mu, q, n^{(i)} \Bigr)
 \equiv \pa_\mu f \frac{\dd x^\mu}{\dd \lambda}
 + \frac{\pa f}{\pa \ln q} \frac{\dd \ln q}{\dd \lambda}
 + D_{(i)} f \frac{\dd n^{(i)}}{\dd \lambda}\,.
 \end{equation}
In a similar manner, one can also decompose the collision term, that
is the r.h.s of Eq.~(\ref{Evolfmunu}) as
\begin{gather}
 C_{\mu \nu} \equiv \frac{1}{2} C^I S_{\mu \nu} + C^P_{\mu \nu}
 + \frac{\ii}{2} C^V \epsilon_{\rho \mu \nu \sigma} e_{(0)}{}^\rho n^\sigma \,,
 \label{PolC}
\end{gather}
so that after extracting the trace, symmetric traceless and
antisymmetric parts we get obviously $L[I]=C^I$, $L[{\bf P}]_{\mu\nu}
= C^P_{\mu\nu}$ and  $L[V]=C^V$. Beware that this does not mean that
the intensity, linear polarization and circular polarization evolve
independently, since for instance $C^I$ is the "intensity part" of the
collision term but it may involve in general all components $I$,
$P_{\mu\nu}$ and $V$ of the tensor-valued distribution function. As a
matter of fact, Compton collision does indeed intermix intensity and
linear polarization, whereas circular polarization evolves
independently.

\subsection{Geodesic equation and momentum evolution}

From Eq (\ref{p0pi}) and the definition of momentum,
 $\dd x^\mu/\dd \lambda=p^\mu$ we obtain
\begin{align}
 \frac{\dd \eta}{ \dd \lambda} &= \frac{q}{a^2} (1 - \alpha) \,,
\label{eta-evo} \\
 \frac{\dd x^i}{ \dd \lambda} &= \frac{q}{a^2} \left( n^{(i)} - \beta^i
 - {h^i}_j  n^{(j)} \right) \,,
\label{xi-evo}
\end{align}
where only the first order terms are kept. 
In terms of $(q, n^{(i)})$, the geodesic equation leads to
 the evolution equation for the conformal energy
\begin{align}
 \frac{\dd \ln q}{\dd \lambda}
 &= \frac{q}{a^2} \Bigl[ - \alpha_{,i} n^{(i)} + \beta_{i, j} n^{(i)} n^{(j)}
 - h_{ij}{}' n^{(i)} n^{(j)} + \alpha \Bigl( \alpha_{,i} n^{(i)}
 - \beta_{i, j} n^{(i)} n^{(j)} + h_{ij}{}'  n^{(i)} n^{(j)} \Bigr) \notag\\
 & \qquad \qquad
 + \alpha_{, j} h^j{}_i n^{(i)} + \beta^k h_{i j, k} n^{(i)} n^{(j)}
 + (\beta_{k, i} - \beta_{i ,k} + 2 h_{i k}{}') h^k{}_j n^{(i)} n^{(j)} \Bigr]
 \,.
\label{q-evo}
\end{align}
As for the direction evolution, up to first order in perturbations, we obtain
\begin{equation}
 \frac{\dd n^{(i)}}{\dd \lambda}
 = -\frac{q}{a^2} S^{(i) (j)} \Bigl[ \alpha_{,j} - (\beta_{k ,j}
 - h_{j k}{}') n^{(k)} + (h_{j l,k} - h_{k l,j}) n^{(k)}n^{(l)} \Bigr] \,.
\label{n-evo}
\end{equation}
It is obvious that $n_{(i)}\dd n^{(i)}/\dd \lambda=1$ as it ought to be
since $n^{(i)}$ is a unit vector.
We need these expressions only at first order, except for the
evolution of $q$ because the background distribution function is
constant in space-time (see below).

Before closing this subsection, we mention that we are free to choose
another affine parameter than $\lambda$, to label a point on a
geodesic. A convenient choice is to take the conformal time $\eta$ at each point of space-time
crossed by the geodesic. The advantage of such choice, is that for
photons having the same direction (and which thus follow the same path), but not the same energy, the same
conformal time $\eta$ would correspond to the same point of the geodesic.
We can trade $\lambda$ for $\eta$ using Eq (\ref{eta-evo}), that is with
\begin{align}
 \frac{\dd \eta}{ \dd \lambda} = \frac{q}{a^2} (1 - \alpha) \quad
 & \Longrightarrow \quad
 \frac{\dd \lambda}{ \dd \eta} = \frac{a^2}{q} (1 + \alpha ) \,.
\end{align}
The evolution of position, conformal energy, and direction, take then
the form
\begin{subequations}
\begin{align}
 \frac{\dd x^i}{\dd \eta}
 = \frac{\dd x^i}{ \dd \lambda} \frac{\dd \lambda}{\dd \eta}
 &= n^{(i)} + \alpha n^{(i)} - \beta^i - {h^i}_j n^{(j)} \,, \\
 \frac{\dd \ln q}{\dd \eta}
 = \frac{\dd \ln q}{ \dd \lambda} \frac{\dd \lambda}{\dd \eta}
 &= - \alpha_{,i} n^{(i)} + \beta_{i, j} n^{(i)} n^{(j)}
 - h_{ij}{}' n^{(i)} n^{(j)} \notag\\
 & \qquad
 + \alpha_{, j} h^j{}_i n^{(i)} + \beta^k h_{i j, k} n^{(i)} n^{(j)}
 + (\beta_{k, i} - \beta_{i ,k} + 2 h_{i k}{}') h^k{}_j n^{(i)} n^{(j)} \,, \\
 \frac{\dd n^{(i)}}{\dd \eta}
 = \frac{\dd n^{(i)}}{ \dd \lambda} \frac{\dd \lambda}{\dd \eta}
 &= - S^{(i) (j)} \left[ \alpha_{,j} - (\beta_{k ,j} - h_{j k}{}') n^{(k)}
 + (h_{j l,k} - h_{k l,j}) n^{(k)}n^{(l)} \right] \,.
\end{align}
\end{subequations}

\subsection{Spectral decomposition of the Boltzmann equation}

We are now in position of writing down explicitly the Boltzmann
equation, expanding the orders of perturbations, and separating the
spectral components. Let us first look at the formal structure of the Boltzmann equation,
especially focusing on the spectral decomposition. At the background level, the Boltzmann equation yields
\begin{equation}\label{EqbackgroundI}
 \frac{\calD}{\calD \lambda} \BBdisB
 = \frac{q}{a^2} \left. \frac{\pa \BBdis (x)}{\pa \eta}
 \right|_{q}
 = - \frac{\dd \ln (a \bar T)}{\dd \lambda}
 \left.\frac{\dd \BBdis(x)}{\dd \ln x} \right|_{x=q/(a\bar T)}
 = 0 \,.
\end{equation}
This implies that $\bar I$ has no time dependence and $\bar T$ scales as $1/a$.
One can conclude that the Planck distribution does not change in time
if the initial distribution is given by the Planck one. This does not mean that the radiation is not losing energy as the universe expands.
Indeed, since the physical energy of a photon is not the conformal energy $q$
but $q/a$, then $\bar \rho \propto \int \bar I (q/a)^3 \dd q/a \propto a^{-4}$
as expected. This background result for the scaling of $\bar T$ is
useful as it implies that only the partial derivative with respect to
$q$ on $\BBdis[q/(a \bar T)]$ are relevant, and this motivates our use
of the conformal energy.

Now the action of the Liouville operator on the intensity,
Eq~(\ref{intensity}), is expanded up to the second order as
\begin{align}
 L[I] &= \frac{\dd \ln q}{\dd \lambda} \BBparq
 - L \Bigl[ \Theta + \Theta^2 \Bigr] \BBparq
 - \Theta \frac{\dd \ln q}{\dd \lambda} \frac{\pa^2 \BBdis}{\pa \ln q^2}
 + L \left[ y + \frac{1}{2} \Theta^2 \right] \calD_q^2 \BBdis \notag\\
 &= - \left[ L [\Theta] - \frac{\dd \ln q}{\dd \lambda}
 - \Theta \frac{\dd \ln q}{\dd \lambda}
 + 2 \Theta \left( L [\Theta] - \frac{\dd \ln q}{\dd \lambda} \right)
 \right] \BBparq
 + \left[ L [y] + \Theta \left( L [\Theta]
 - \frac{\dd \ln q}{\dd \lambda} \right) \right] \calD^2_q \BBdis \,,
\label{original}
\end{align}
where we used $\pa \BBdis/\pa \eta = 0$.

When we compare this formulation of the Liouville operator with the spectral decomposition~(\ref{intensity}), we
are tempted to say that the expression inside the first square brackets contributes to
 the evolution of the temperature due to
 free-streaming, and that the expression inside the second square
 brackets is very closely related to the evolution of the
 distortion. It order to give a clear meaning to this assertion we
 decompose Eq.~(\ref{original}) according to
\begin{equation}\label{spectraldecI}
 L[I] \equiv \frac{q}{a^2} \left[ - \Bigl( \calL^\Theta + 2 \Theta \calL^\Theta \Bigr) \BBparq
 + \Bigl( \calL^Y + \Theta \calL^\Theta \Bigr) \calD^2_q \BBdis \right] \,.
\end{equation}
This spectral decomposition is motivated by the fact that
 i) in the case where the conformal energy is not affected by free-streaming,
  $\dd \ln q / \dd \lambda = 0$, then $(q/a^2) \calL^\Theta$ simply reduces to $L[\Theta]$;
   and ii) the prefactor $q/a^2$ is introduced because the Liouville term is expected to be proportional to $q/a^2$, as it can be inferred from the explicit form~(\ref{eta-evo}) of $\dd \eta/\dd \lambda$. From a comparison of Eq~(\ref{original}) with this decomposition, we have
\begin{align}\label{SpectraldecLI}
\frac{q}{a^2}{\cal L}^\Theta&=L[\Theta] -(1+ \Theta )\frac{\dd \ln q}{\dd \lambda}\,,\\
 \frac{q}{a^2}{\cal L}^Y&=L[y]\,.
\end{align}
We must bear in mind that in these expressions, even though
the operator $L[.]$ has been defined in Eq.~(\ref{defofLs}) for
functions of $(\eta,x^i,q,n^{(i)})$, it is applied on the spectral
components $\Theta$ and $y$ which do not depend on $q$.

Since the Liouville operator is equated to the collision term in the
Boltzmann equation, it is convenient to decompose the collision term in the same manner
as the Liouville term. That is, it is decomposed as
\begin{equation}
 C^I \equiv \frac{q}{a^2} \left[ - \Bigl( \calC^\Theta + 2 \Theta \calC^\Theta \Bigr) \BBparq
 + \Bigl( \calC^Y + \Theta \calC^\Theta \Bigr) \calD^2_q \BBdis \right] \,,
\end{equation}
such that the spectral components of the Boltzmann equation can be
formally very simple and are given by
 \begin{align}
 \calL^\Theta= \calC^\Theta \,,\qquad \calL^Y=\calC^Y.
 \end{align}
This decomposition means that once the spectral decomposition of the
collision term is known (${\cal C}^\Theta$ and ${\cal C}^Y$), then we only need
to obtain the spectral decomposition of the Liouville term from
Eqs~(\ref{SpectraldecLI}).

We follow the same logic for polarization. First, the corresponding
Liouville operator reads
\begin{align}
 L [ {\bf P} ]_{\mu \nu}
 &= - L \Bigl[ (1 + 3 \Theta) \calP_{\mu\nu} \Bigr] \BBparq
 - \calP_{\mu \nu} \frac{\dd \ln q}{\dd \lambda} \frac{\pa^2 \BBdis}{\pa \ln q^2}
 + L \Bigl[ Y_{\mu \nu} + \Theta \calP_{\mu \nu} \Bigr] \calD_q^2 \BBdis \notag\\
 &= - \left[ (1 + 3 \Theta) L [ {\bm \calP} ]_{\mu \nu}
 + 3 \left(L[\Theta]- \frac{\dd \ln q}{\dd \lambda}\right) \calP_{\mu \nu} \right] \BBparq
 + \left[ L [ {\bf Y} ]_{\mu \nu} + \left(L[\Theta]- \frac{\dd \ln q}{\dd \lambda}\right)\calP_{\mu \nu}
 + \Theta L [{\bm \calP}]_{\mu \nu} \right] \calD_q^2 \BBdis \,. 
\end{align}
For the same reasons as in the case of intensity, it appears natural
to decompose this Liouville operator into spectral components according to
\begin{equation}\label{spdec:calP_lhs}
 L [ {\bf P} ]_{\mu \nu}\equiv\frac{q}{a^2} \left\{ - \Bigl[ (1 + 3 \Theta) \calL_{\mu \nu}^P
 + 3 \calL^\Theta \calP_{\mu \nu} \Bigr] \BBparq
 + \Bigl( \calL^Y_{\mu \nu} + \calL^\Theta \calP_{\mu \nu}
 + \Theta \calL_{\mu \nu}^P \Bigr) \calD_q^2 \BBdis \right\} \,,
\end{equation}
which implies that the spectral components are given by
\begin{equation}\label{SpectraldecLP}
\frac{q}{a^2} \calL_{\mu \nu}^P \equiv L [ {\bm \calP} ]_{\mu \nu},\qquad \frac{q}{a^2} \calL_{\mu \nu}^Y \equiv L [ {\bf Y} ]_{\mu \nu} \,.
\end{equation}
The collision term must then follow the same type of decomposition, that is
\begin{align}
 C^P_{\mu \nu}
 &= \frac{q}{a^2} \left\{ - \Bigl[ (1 + 3 \Theta) \calC_{\mu \nu}^P
 + 3 \calC^\Theta \calP_{\mu \nu} \Bigr] \BBparq
 + \Bigl( \calC^Y_{\mu \nu} + \calC^\Theta \calP_{\mu \nu}
 + \Theta \calC_{\mu \nu}^P \Bigr) \calD_q^2 \BBdis \right\} \,,
\label{spdec:calP_rhs}
\end{align}
so that, again, the spectral components of the polarized part of the Boltzmann equation take the formally simple form
\begin{equation} \label{eq:Boltz_pol_dis}
\calL^P_{\mu \nu}=\calC^P_{\mu \nu} \,,\qquad \calL^Y_{\mu \nu} =\calC^Y_{\mu \nu} \,.
 \end{equation}
Again, this decomposition means that once the spectral decomposition of the
collision term is known (${\cal C}^P_{\mu\nu}$ and ${\cal C}^Y_{\mu\nu}$), then we only need
to obtain the spectral decomposition of the Liouville term from
Eqs~(\ref{SpectraldecLP}) bearing in mind that the Liouville operator
$L[.]$  applies on functions which do not depend on $q$, but only on $(\eta,x^i,n^{(i)})$.

\subsection{Temperature and spectral distortion of Liouville operators}

Now that the spectral separation of the Boltzmann equation is performed, it is time to expand
the equations obtained in orders of perturbations. In the next two
sections, we present such expansion for the temperature and spectral
distortion parts of the Boltzmann equation. The case of polarization
is reported in Appendix~\ref{sapp:Boltzeq_pol}.

At first order in perturbation, with Eqs.~(\ref{eta-evo}) and (\ref{q-evo}),
the Boltzmann equation leads to
\begin{align}
 {\cal L}^\Theta&= \Theta' + \Theta_{, i} n^{(i)}
 + \alpha_{, i} n^{(i)} - \beta_{i, j} n^{(i)} n^{(j)}
 + h_{i j}{}' n^{(i)} n^{(j)} \,.
\end{align}
Note that there is absolutely no q-dependence, nor scale factor $a$ in this expression,
meaning that our spectral decomposition performed in Eq.~(\ref{spectraldecI}) is adequate. Concerning the
spectral distortion part, the Liouville part at first order is ${\cal L}^Y= y'+y_{,i}
n^{(i)}$, but since ${\cal C}^Y=0$ at first order (see
section~\ref{ssec:Collision}), one can conclude that only the temperature part evolves at first order
and no spectral distortion is induced.

Up to the second order, the Boltzmann equation for the temperature is given by
\begin{align}\label{LCThetanotOpened}
 {\cal L}^\Theta
 &= \frac{a^2}{q}\left[L[\Theta]
 - (1 + \Theta) \frac{\dd \ln q}{\dd \lambda}\right] \notag\\
 &= \Theta' + \Theta_{, i} n^{(i)}
 + \alpha_{, i} n^{(i)} - \beta_{i, j} n^{(i)} n^{(j)}
 + h_{i j}{}' n^{(i)} n^{(j)} \notag\\
 & \qquad 
 +\frac{a^2}{q}
 \left[ \left. \frac{\dd \eta}{\dd \lambda} \right|^{(1)} \Theta'
 + \left. \frac{\dd x^i}{\dd \lambda} \right|^{(1)} \Theta_{, i}
 + \left. \frac{\dd n^{(i)}}{\dd \lambda} \right|^{(1)} D_i \Theta
 - \left. \frac{\dd \ln q}{\dd \lambda} \right|^{(1) \times (1)}
 - \Theta \left. \frac{\dd \ln q}{\dd \lambda} \right|^{(1)} \right] 
  = \calC^\Theta \,.
\end{align}
As for the spectral distortion, the Boltzmann equation is given, even at second order by
\begin{align}
 {\cal L}^Y &= y' + y_{, i} n^{(i)} = \calC^Y \,.
\end{align}
This means simply that gravitational effects do not induce spectral
distortions, and this result holds actually  non-perturbatively.

Before ending this subsection, for the sake of completeness, we shall
write down the most general form of the second order Boltzmann
equation for the intensity. Using Eqs.~(\ref{eta-evo}), (\ref{q-evo}),
(\ref{xi-evo}) and (\ref{n-evo}), the detailed  form  of the evolution
equation for temperature obtained in Eq.~(\ref{LCThetanotOpened}) is
\begin{align}
 {\cal L}^\Theta=& \Theta' + \Theta_{, i} n^{(i)} + \alpha_{,i} n^{(i)}
 - \beta_{i ,j} n^{(i)} n^{(j)} + h_{i j}{}' n^{(i)} n^{(j)} \notag\\
 & \quad
 - \alpha \Theta' - (\beta^i + h^i{}_j n^{(j)}) \Theta_{, i}
 - \Bigl[ \alpha_{,i} - (\beta_{j ,i} - h_{i j}{}') n^{(j)}
 + (h_{k i, j} - h_{j k, i}) n^{(j)} n^{(k)} \Bigr] D^i \Theta \notag\\
 & \quad
 - \Bigl( \alpha \alpha_{,i} + \alpha_{, j} h^j{}_i \Bigr) n^{(i)}
 - \Bigl[ \alpha (- \beta_{i ,j} + h_{i j}{}') + \beta^k h_{i j, k}
 + (\beta_{k, i} - \beta_{i, k} + 2 h_{i k}{}') h^k{}_j \Bigr] n^{(i)} n^{(j)}
 \notag\\
 & \quad
 - \Theta n^{(i)} \Bigl( - \alpha_{,i} + \beta_{i ,j} n^{(j)} - h_{i j}{}' n^{(j)} \Bigr) 
 = \calC^\Theta \,.
\end{align}

\subsection{Temperature and spectral distortion of collision terms}
\label{ssec:Collision}

The expression of the collision term has been derived by taking into account
only the intensity in~\cite{Dodelson1993,Bartolo:2006cu} and then it was extended
to include the effect of polarisation in~\cite{Pitrou:2008hy,Pitrou:2008ut,Beneke:2010eg}.
Here we summarize the result obtained by Beneke et
al. \cite{Beneke:2010eg} applying the decomposition of the distribution function into intensity and linear polarisation.

The complete expression of the collision term for intensity is given by
\begin{align}
 \calC^\Theta
 &= a \, \bar{n}_e \sigma_T \left( - \Theta + \langle \Theta \rangle
 - \frac{3}{4} S^{(i) (j)} \Bigl[ \langle \Theta m_{(i) (j)} \rangle
 - 2 \langle \calP_{(i) (j)} \rangle \Bigr]
 + v^{(i)} n_{(i)} + \calS^T + S^{(i) (j)} \calQ^T_{(i) (j)}
  + \delta_e \calC^\Theta \right) \,, \\
 \calC^Y &= a \, \bar{n}_e \sigma_T \left( - y + \langle y \rangle
 - \frac{3}{4} S^{(i) (j)} \Bigl[ \langle y m_{(i) (j)} \rangle
 - 2 \langle Y_{(i) (j)} \rangle \Bigr]
 + \calS^Y + S^{(i) (j)} \calQ^Y_{(i) (j)} \right) \,,
\end{align}
where $\calS^T$, $\calS^Y$,
$\calQ^T_{(i) (j)}$ and $\calQ^Y_{(i) (j)}$ are quadratic
contributions defined by
 \begin{align}
 \calS^T
 &= \Theta^2 + \langle \Theta^2 \rangle 
  - 2 \Theta \langle \Theta \rangle - \Theta v^{(i)} n_{(i)}
  + 2 \langle \Theta \rangle v^{(i)} n_{(i)} 
  - 2  \langle \Theta n_{(i)} \rangle v^{(i)} - \frac{1}{5} v^{(i)} v_{(i)}
 + (v^{(i)} n_{(i)})^2 \,, \\
 \calS^Y
 &= \frac{1}{2} \Theta^2 +  \frac{1}{2} \langle \Theta^2 \rangle
 - \Theta \langle \Theta \rangle - \Theta v^{(i)} n_{(i)} 
 + \langle \Theta \rangle v^{(i)} n_{(i)}
 - \langle \Theta n_{(i)} \rangle v^{(i)} 
 + \frac{1}{5} v^{(i)} v_{(i)} + \frac{1}{2} (v^{(i)} n_{(i)})^2 \,,
 \end{align}
\begin{align}
 \calQ^T_{(i) (j)}
 &= - \frac{3}{4} \langle \Theta^2 m_{(i) (j)} \rangle
 + \frac{3}{2} \Theta \langle \Theta m_{(i) (j)} \rangle
 + \frac{3}{4} \Bigl[ \langle \Theta n_{(j)} \rangle v_{(i)}
 + \langle \Theta n_{(i)} \rangle v_{(j)} \Bigr] \notag\\
 & \qquad
 - \frac{3}{4} n^{(k)} \Bigl[ v_{(j)} \langle \Theta m_{(i) (k)} \rangle
 + v_{(i)} \langle \Theta m_{(j) (k)} \rangle \Bigr]
 - \frac{3}{4} v^{(k)} \Bigl[ 2 n_{(k)} \langle \Theta m_{(i) (j)} \rangle
 + \langle \Theta m_{(i) (j)} n_{(k)} \rangle \Bigr]
 - \frac{1}{5} v_{(i)} v_{(j)} \notag\\
 & \qquad
 + \frac{9}{2} \langle \Theta \calP_{(i) (j)} \rangle
 - 3 \Theta \langle \calP_{(i) (j)} \rangle
 + \frac{3}{2} n^{(k)} \Bigl[ v_{(j)} \langle \calP_{(i) (k)} \rangle
 + v_{(i)} \langle \calP_{(j) (k)} \rangle \Bigr] \notag\\
 & \qquad
 + \frac{3}{2} v^{(k)} \Bigl[ \langle \calP_{(i) (k)} n_{(j)} \rangle
 + \langle \calP_{(j) (k)} n_{(i)} \rangle \Bigr]
 - \frac{3}{2} v^{(k)} \Bigl[ \langle \calP_{(i) (j)} n_{(k)} \rangle
 - 2 n_{(k)} \langle \calP_{(i) (j)} \rangle \Bigr] \,, \\
 \calQ^Y_{(i) (j)}
 &= - \frac{3}{8} \langle \Theta^2 m_{(i) (j)} \rangle
 + \frac{3}{4} \Theta \langle \Theta m_{(i) (j)} \rangle
 - \frac{3}{4} v^{(k)} \Bigl[ n_{(k)} \langle \Theta m_{(i) (j)} \rangle
 - \langle \Theta m_{(i) (j)} n_{(k)} \rangle \Bigr]
 - \frac{1}{20} v_{(i)} v_{(j)} \notag\\
 & \qquad
 + \frac{3}{2} \langle \Theta \calP_{(i) (j)} \rangle
 - \frac{3}{2} \Theta \langle \calP_{(i) (j)} \rangle
 + \frac{3}{2} v^{(k)} \Bigl[ n_{(k)} \langle \calP_{(i) (j)} \rangle
 - \langle \calP_{(i) (j)} n_{(k)} \rangle \Bigr] \,,
\end{align}
 and $m_{(i) (j)} \equiv n_{(i)} n_{(j)} - \delta_{(i) (j)}/3$. Note
 that we have introduced the notation
 $\langle Q \rangle = \int_{\Omega} Q = \int \dd^2 n^{(i)} Q$, which
 corresponds to a multipole extraction that we do not perform
 explicitly here.  Note also that $\delta_e=\delta n_e/\bar n_e$ is
 the fractional perturbation of the baryons number density, and
 $v^{(i)}$ are the tetrad components of the baryons spatial velocity.
Again, we also defer the expression of the collision term for polarisation
to Appendix~\ref{sapp:Boltzeq_pol}.

\section{Gauge dependence of the distribution function}
\label{sec:gauge}

Now that we have established the Boltzmann equation, up to second
order, with its spectral components separated, we investigate the
gauge dependence of its constituents. Eventually the Boltzmann
equation itself should be gauge-invariant, so if we are able to check
explicitly that the Boltzmann equation is gauge-invariant, this means
that it is very likely that i) the perturbative expansion of the
equation is correct; and ii) the gauge transformation rules for all
its constituents (metric and distribution function perturbations) are correctly understood. We thus consider this
verification as a \emph{consistency test}. This section is dedicated
entirely to this task. We first review the gauge dependence
for tensors, and deduce how it can be extended to a scalar
distribution function. The case of a tensor-valued distribution
function, even though it is the less trivial part, is treated in
appendix~\ref{AppGTtensor}. We then infer what should be the transformation rule of
the Liouville and collision operators, and in order to complete the
consistency test, we check that the perturbed expressions of the
Liouville and collision operators do indeed transform following
these rules.

\subsection{Coordinates on the manifold}

We need to specify how the functional dependence of the quantities appearing in the Boltzmann equation (and in the Einstein equation) is obtained.
If we consider a scalar function $f:{\cal M}\mapsto \RR$ on the space-time manifold ${\cal M}$, then the choice of a coordinates system \footnote{We assume for the simplicity of the argument that one system of coordinates is enough to cover the entire manifold.}
$c:\RR^4 \mapsto {\cal M}$ does not affect the geometrical meaning of
this function, but it affects its functional form $f \circ c:\RR^4 \mapsto \RR$, where $\circ$ designates the composition rule, in the sense that for another coordinates system
 $\tilde c:\RR^4 \mapsto {\cal M}$, then $f\circ \tilde c \neq f \circ c$.
The confusion only arises from the fact that we often refer to $f \circ c$ as $f$ only.
A distribution function $f$ (that we take as a scalar-valued for simplicity here)
is a function on the tangent bundle $T {\cal M}$ of the manifold and can be regarded as
a function on manifold ${\cal M}$ describing the space-time and on the tangent space $\RR^4$ (more precisely a restriction to the mass shell $\RR^3$) of each point.
It is thus a function
\begin{equation}
 f: (T{\cal M})\mapsto \RR\,.
\end{equation}
Again the particular choice of coordinates on the tangent space does not affect the geometrical meaning of the function but its functional form.
Once a choice $c$ of coordinates on the manifold ${\cal M}$ has been made,
there is a natural basis, called canonical basis, that is made of the partial derivatives with respect to the coordinates. This leads to a natural coordinates system $Tc$ for the tangent space at each point. Thus $(c,Tc) : \RR^4 \times \RR^3 \mapsto T{\cal M}$,
is a coordinates system for the tangent bundle. In order to simplify the notation we will note $(c,Tc)$ as simply as $c$.

Furthermore, in this paper we use coordinates in the tangent space described
by the tetrad components. More specifically we use the components of the conformal
momentum in spherical coordinates in this tetrad basis, $(q,n^{(i)})$.
There is a problem with such a choice since the tetrad basis is not unique.
However, once a choice $c$ of coordinates on the manifold ${\cal M}$
has been made, the tetrad basis might be completely fixed from
the metric through a prescription described in \S~\ref{ssec:deftet}.
Once a coordinates system $c$ has been chosen, and the tetrad is fixed
thanks to this choice, we obtain the functional form of $f$ in the form $f \circ c:\RR^4\times \RR^3\mapsto \RR$.

\subsection{Geometrical interpretation of the gauge}

When performing perturbations around a background FLRW space-time,
we need to have a one-to-one correspondence between the background space-time $\overline{\cal M}$ and the physical (and perturbed) space-time ${\cal M}$.
This can be completely defined geometrically~\cite{Bruni:1996im}
but we take a shorter approach. If we have two sets of coordinates
 \footnote{Again here for simplicity, we
assume that such coordinates system covers the whole manifold.}
 $\bar c:\RR^4 \mapsto \overline{\cal M}$ and $c:\RR^4 \mapsto {\cal M}$
on the background and the perturbed space-time, then we identify points with the
same coordinates, that is, we identify points with $c \circ \bar c^{-1}:\overline{\cal M}\mapsto {\cal M}$.  Since we could have chosen different sets of coordinates,
there is some freedom in this choice, which is known as the gauge freedom.

\begin{figure}[htb]
\begin{center}
	\includegraphics[width=\textwidth]{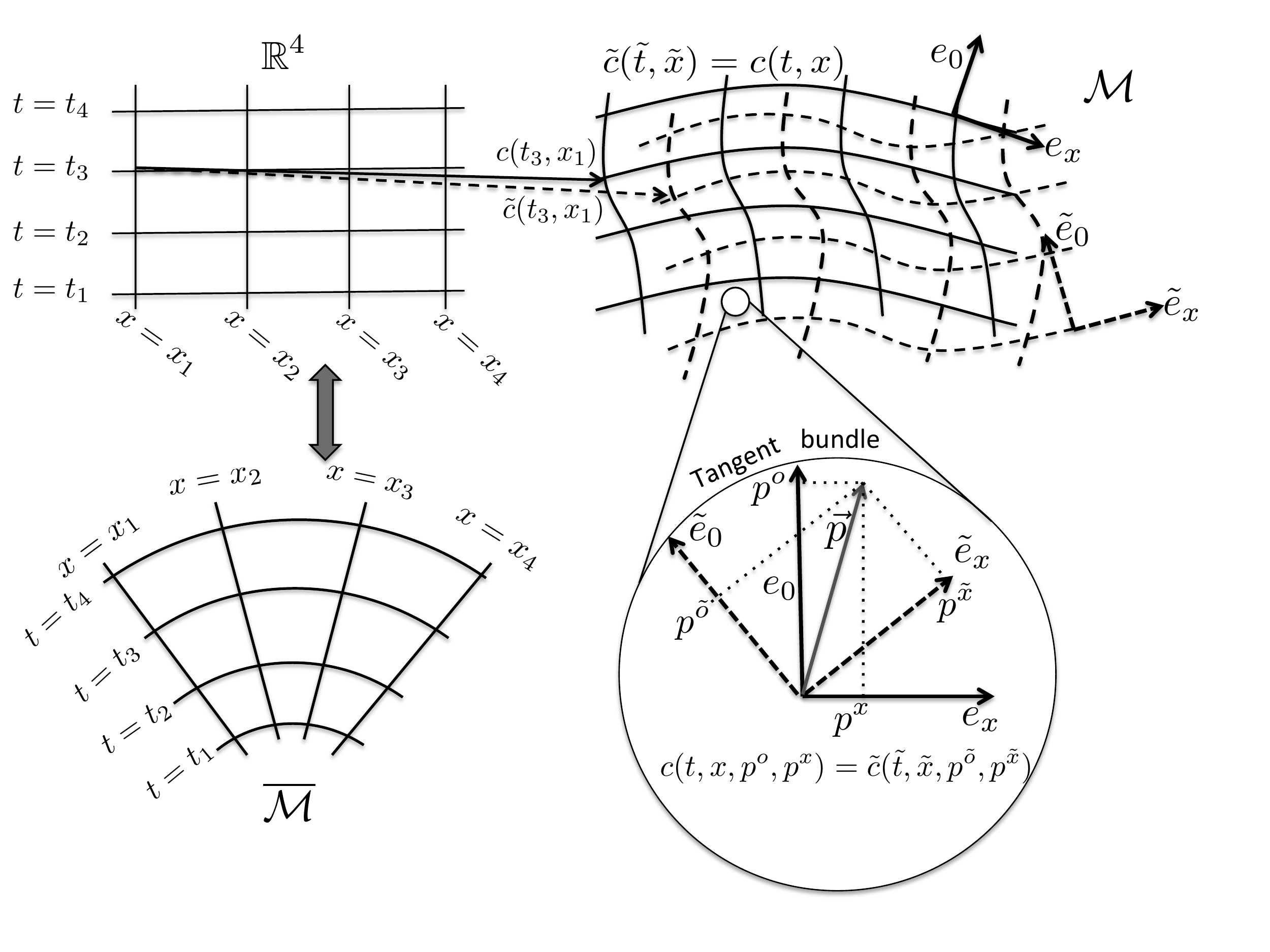}
	\caption{\label{fig1} In this figure, we represent all four dimensional
          spacetimes with only two dimensions. First, we have noted that there is essentially a unique way to map
          $\RR^4$ to the background manifold, that is to relate the
          top-left to the bottom left. However, there are several ways
          to relate $\RR^4$ to the perturbed manifold, and
          consequently to relate the background manifold to the
          perturbed manifold. We represented a coordinates system $c$
          and another coordinates system $\tilde c$ which relate
          $\RR^4$ (top-left) to the physical manifold (top-right). For
          each point $P$ of the physical manifold, there is a set of four
          numbers $x^\mu$ and another set of four numbers $\tilde
          x^\mu$ such that $c(x^\mu) = \tilde c(\tilde
          x^\mu)$. Furthermore, each coordinates system has different
          surfaces of constant time, and thus different set of
          tetrads, given that the null tetrad is always chosen to be
          orthogonal to the constant-time surfaces. When using
          the tetrad field to extract the components of a given
          momentum (that is a point in the tangent bundle), this will
          lead to different components, depending on the coordinates
          system chosen, but since this is the same momentum at the
          same point and the tetrad are normalized, we can relate the
          components by a Lorentz transformation. For a given point of the tangent
          bundle, that is, for a given point of the manifold and a
          given momentum, we always have
          $c(x^\mu,p^{(a)})=\tilde c(\tilde x^\mu, p^{\tilde{(a)}})$.}
\end{center}
\end{figure}

On the background, the symmetries can justify that we can find
a preferred choice of coordinates. For instance for a flat FLRW space-time
within a given background cosmology, it is enough to choose that the
time coordinate is the proper time of observers with 4-velocity
orthogonal to the homogeneous surfaces, that is the proper time of comoving
observers. On a given homogeneous surface, there are preferred choices
of Cartesian coordinates, since it is conformally related to $\RR^3$,
and all these Cartesian systems on the spatial homogeneous surfaces are related by global translation and rotation in $\RR^3$ which
are irrelevant given the homogeneity. And once a coordinate system has been chosen on a
homogeneous surface it can be Lie dragged by the comoving observers to any
homogeneous surface. So essentially there is a unique mapping $\bar c$
from $\RR^4$ to the background manifold $\overline{\cal M}$. This
point is illustrated in the left part of Fig.~\ref{fig1}.

However on the physical space-time we could consider another
coordinates system $\tilde c:\RR^4 \mapsto {\cal M}$, and
this leads to a different identification through $\tilde c \circ \bar c^{-1}$.
The fact that we fix the system of coordinates on the background space-time
but there is still some freedom on the physical space-time
for the choice of coordinates leads to a freedom
in the identification between points of these two space-times.
With $c$, a given point $P \in {\cal M}$ would be labelled
by the coordinates $x^\mu$, that is $c( {\bm x}) = P$,
and with $\tilde c$ it would be labeled by the coordinates $\tilde x^\mu$,
that is $\tilde c( \tilde{\bm x}) = P$ (see the upper part of Fig.~\ref{fig1}).
For every point, there exist four numbers ${\xi}^\mu= ({ T},{ L}^i)$ such that
\begin{gather}\label{Trulexmu}
 \tilde{x}^\mu( {\bm x}) = x^\mu + {\xi}^\mu( {\bm x}) \,.
\end{gather}
We should note that, although there is an index in the notation,
${\xi}^\mu$ is not a vector field on ${\cal M}$ nor
on $\overline{\cal M}$, but it can be seen as a vector field on $\RR^4$.
In the literature, a vector field $\zeta^\nu$ is often used to generate a coordinate transformation ~\cite{Bruni:1996im}
 \begin{equation}
 \tilde x^\mu = \exp\left({\cal L}_{\bzeta}\right)x^\mu
 =x^\mu+\zeta^\mu+\frac{1}{2}\zeta^\mu{}_{,\nu} \zeta^\nu+\cdots\,.
\end{equation}
In this paper we adopt the former definition Eq.~(\ref{Trulexmu}) and care must be taken when comparing our transformation rules with those in the literature.
Note that in the rest of this section, we will use extensively the notation
\begin{equation}
\tilde{\bm x} \equiv \tilde x^\mu\,,\qquad {\bm x}\equiv x^\mu\,,
\end{equation}
even though $x^\mu$ and $\tilde x^\mu$ are not vectors but just coordinates.


\subsection{Metric transformation}
In general the coordinate transformation rule of any tensorial quantity is
 given by
\begin{equation}\label{GTTensor}
 {\bm T} \circ \tilde c{(\tilde {\bm x})} = {\bm T}\circ c({\bm x})\,.
\end{equation}
This means that it is invariant under a coordinates transformation since it is geometrically defined and independent of the coordinates used to parametrise
the manifold. From this we can deduce the gauge transformation of its components,
which is defined as the transformation when the tensors are compared at the same coordinate in two different coordinates systems.
Let us consider in particular the metric.
Since $g_{\mu \nu} \equiv {\bm g}(\pa/\pa x^\mu,\pa/\pa x^\nu)$ and
 $g_{\tilde \mu \tilde \nu} \equiv
 {\bm g}(\pa/\pa \tilde x^\mu,\pa/\pa \tilde x^\nu)$,
 we then deduce the usual coordinate transformation rule of a 2-form
\begin{equation}\label{Transfog1}
g_{\tilde \mu \tilde \nu} \circ \tilde c{(\tilde {\bm x})} = \frac{\pa
  x^\alpha}{\pa \tilde x^\mu} \frac{\pa
  x^\beta}{\pa \tilde x^\nu}  g_{\alpha \beta}\circ c({\bm x})\,.
\end{equation}
Expanding the left hand side which is evaluated at $\tilde {\bm x}$
around ${\bm x}$ leads the gauge transformation rule up to the second order
\begin{gather}\label{EqTmetric}
g_{\tilde \alpha \tilde \beta}\circ \tilde c{\left({\bm x}\right)} =
g_{\alpha \beta}\circ c({\bm x})
 - \calL_{{\bxi}} g_{\alpha \beta}\circ c({\bm x}) + \frac{1}{2}
 \calL^2_{{\bxi}} g_{\alpha \beta}\circ c({\bm x})
 + \frac{1}{2} \calL_{{\bxi} {\bxi}} g_{\alpha \beta}\circ c({\bm x}) \,,
\end{gather}
where ${\bxi} {\bxi}$ designates the quantity ${\xi}^\mu{}_{,\nu}{\xi}^\nu$.
We emphasize that the coordinate transformation~\eqref{Transfog1}
is the transformation of components at the same point of space-time,
whereas the gauge transformation~\eqref{EqTmetric} is
the transformation of the components at different points which have the same coordinates
in two different coordinates systems.
Since the notation can become rather cumbersome if we specify which coordinates system $c$ or $\tilde c$ is to be used, we shall indicate it only when it is the new coordinates system $\tilde c$.

Throughout this paper, we use the symmetrization and anti-symmetrization definitions
\begin{equation}
 X_{(\mu\nu)} \equiv \frac{1}{2}\left(X_{\mu\nu}+ X_{\nu\mu}\right) \,, \qquad
 X_{[\mu\nu]} \equiv \frac{1}{2}\left(X_{\mu\nu}-X_{\nu\mu}\right)\,.
\end{equation}
Up to second order, from the transformation~(\ref{EqTmetric}) and the
decomposition~(\ref{MetricADM}), we obtain the gauge transformation for
the ADM variables as
\begin{subequations}\label{TrulemetricADM}
\begin{align}
 \tilde{\alpha}
 &= \alpha - \calH T - T' + \frac{1}{2} (\calH^2 + \calH') T^2
 + \calH (2 T' T + T_{,i} L^i) - \calH \alpha T \notag\\
 & \qquad - \alpha T' - \alpha' T - \alpha_{,i} L^i + \beta^i T_{,i}
 + T'' T + T'{}^2 + T'_{,i} L^i + \frac{1}{2} T^{, i} T_{, i} \,, \\
 \tilde{\beta}^i
 &= \beta^i + T^{,i} - L^i{}' + 2 \alpha T^{, i} - \beta^i T'
 - \beta^i{}' T + \beta^j L^i{}_{,j} - \beta^i{}_{,j} L^j - 2 h^{i j} T_{,j}
 \notag\\
 & \qquad - (2 T' T^{, i} + T T'{}^{, i}) + T L^i{}'' + T' L^i{}'
 + T_{,j} L^{i ,j} - T^{, i}{}_{,j} L^j + L^i{}_{,j}{}' L^j \,, \\
 2 \tilde{h}_{i j} &= 2 h_{i j} - 2 \calH T \delta_{i j} - 2 L_{(i ,j)}
 - 4 \calH T h_{i j} + (2 \calH^2 + \calH') T^2 \delta_{i j}
 + 2 \calH ( T T' + T_{,k} L^k) \delta_{i j} \notag\\
 & \qquad
 + 4 \calH T L_{(i ,j)} - 2 \beta_{(i} T_{,j)} - 2 h_{i j}{}' T
 - 2 h_{i j, k} L^k - 4 h_{(i| k} L^k{}_{,|j)} \notag\\
 & \qquad
 - T_{,i} T_{,j} + 2 T L_{(i ,j)}{}'
 + 2 T_{(,i} L_{j)}{}' + 2 L_{(i ,j) k} L^k
 + 2 L_{(i|, k} L^k{}_{,|j)} + L_{k, i} L^k{}_{, j} \,.
\end{align}
\end{subequations}
These relations must be understood as follows. $a^2 \alpha$ at first order
is the $00$ component of the first order metric in the $c$
coordinates system, and taken at the point of coordinates ${\bm x}$, and is thus equal to
$-\frac{1}{2}g^{(1)}_{00}\circ c({\bm x})$. Instead, $a^2 \tilde \alpha$ which
means $a^2 \tilde \alpha({\bm x})$, when considered at first order is
the $\tilde 0 \tilde 0$ component of
the first order metric in the $\tilde c$ coordinates system, but also taken at the point
of coordinates ${\bm x}$, that is $-\frac{1}{2}g^{(1)}_{\tilde 0 \tilde
  0}\circ \tilde c{({\bm x})}$.

\subsection{Tangent space basis and tetrads}

As discussed in Section~\ref{ssec:f}, the transformation of the basis on the tangent space is entirely linked to the coordinates change in the base manifold. Usually, the canonical basis $\pa/\pa x^\mu$ and the corresponding forms $\dd x^\mu$ are
used as a basis of the tangent space and they transform according to
\begin{equation}\label{TruleCanonicalBasis}
\frac{\pa}{\pa \tilde x^\mu}\circ \tilde c{(\tilde{\bm x})}
 = \frac{\pa x^\nu}{\pa \tilde x^\mu} \frac{\pa}{\pa  x^\nu}{({\bm x})} \,,
 \qquad \dd \tilde x^\mu \circ \tilde c {(\tilde{\bm x})}
 = \frac{\pa \tilde x^\mu}{\pa x^\nu} \dd x^\nu {c({\bm x})} \,.
\end{equation}
When we consider the components of the metric, these components refer
to this canonical basis. However we will use the tetrad field as a
basis for the tangent space. We thus need to relate $\tilde{\bm
  e}_{(a)}\circ \tilde c{(\tilde{\bm x})}$ with  ${\bm  e}_{(a)}\circ c{({\bm x})}$ in a similar fashion. Since this is a relation between two orthonormal basis at the same point of space-time, there exists a
Lorentz transformation $\Lambda^{(a)}_{~~ (b)}$ such that we can
relate the two tetrad fields associated with the coordinates systems
$c$ and $\tilde c$  (see the bottom right part of Fig.~\ref{fig1} for
an illustration of this) as
\begin{equation}
 \tilde{\bm e}_{(a)}\circ \tilde c{(\tilde{\bm x})}
 = \Lambda_{(a)}^{~~ (b)} ({\bm x}) {\bm e}_{(b)} ({\bm x}) \,,\qquad
  \tilde{\bm e}^{(a)} \circ \tilde c{(\tilde{\bm x})}
 = \Lambda^{(a)}_{~~ (b)} ({\bm x}) {\bm e}^{(b)} ({\bm x})\,,\qquad
 \eta_{(c) (d)} \Lambda^{(c)}_{~~ (a)} \Lambda^{(d)}_{~~ (b)} = \eta_{(a) (b)} \,.
\end{equation}
The components of this Lorentz transformation are given, up to first order, by
\begin{equation}\label{Reltetradssamephysicalpoint}
 \Lambda^{(0)}_{~~ (0)} = 1 \,,\qquad
 \Lambda^{(0)}_{~~ (i)} = \Lambda^{(i)}_{~~ (0)} = \pa_i T \,,\qquad
 \Lambda^{(i)}_{~~ (j)} = \delta^{(i)}_{~~ (j)} +L^{[i}_{~~ , j]}\,.
\end{equation}
At second order, it proves more useful to relate the components of
these quantities, that is, to relate
$\widetilde{e}_{(a)}{}^{\widetilde{\mu}}\circ \tilde c{(\tilde{\bm x})}
 \equiv \dd \tilde x^\mu\circ \tilde c{(\tilde{\bm    x})}
 [\widetilde{\bm e}_{(a)}]$ to $
 e_{(a)}{}^\mu({\bm x})\equiv \dd x^\mu({\bm x})[{\bm e}_{(a)}]$.
For the former components, we must use~\eqref{tetrads}
 with $a(\tilde{\eta})$, $\tilde \alpha(\tilde{\bm x})$,
 $\tilde \beta_i(\tilde{\bm x})$ and $\tilde h_{ij}(\tilde{\bm x})$
 [that is  $g_{\tilde \mu \tilde \nu}\circ \tilde c{(\tilde{\bm x})}$]
 which can be deduced from the rule~\eqref{TrulemetricADM} just by shifting the argument of
 the left hand side of the rules~\eqref{TrulemetricADM} from ${\bm x}$ to
 $\tilde{\bm x}$.
For the latter we must use~\eqref{tetrads} with $a(\eta)$, $\alpha({\bm x})$,
 $\beta_i({\bm x})$ and $h_{ij}({\bm x})$ [that is  $g_{\mu \nu}{({\bm
     x})}$]. We do not report the corresponding expression since we will work instead
 directly with the perturbation components of the metric in the next section.

\subsection{Momentum, energy $q$ and direction $n^{(i)}$}

The components of the momentum of a particle in the canonical basis transform as
\begin{equation}\label{Transfopmu}
 p^{\widetilde{\mu}}\circ \tilde c(\tilde{\bm x})
 = \frac{\pa \tilde{x}^\mu}{\pa x^\nu} p^\nu({\bm x})\,, \qquad
 \text{with} \qquad p^{\widetilde{\mu}} \equiv \dd \tilde
 x^\mu({\bm p})\,,\quad p^{\mu} \equiv \dd x^\mu({\bm p})\,.
\end{equation}
However, we are going to use the tetrad basis in the tangent space
rather than the canonical basis. We thus want to express
$p^{\widetilde{(a)}}\circ \tilde c{(\tilde{\bm x})}
 \equiv \widetilde{\bm e}^{{(a)}}\circ \tilde c{(\tilde{\bm x})}[{\bm p}]$,
 [or $\tilde q\circ \tilde c{(\tilde{ \bm x})}$
 \footnote{Under our conventions, $\tilde q \equiv a p^{\widetilde{(0)}}$.}
 and $n^{\widetilde{(\imath)}}\circ \tilde c{(\tilde{ \bm x})}$]
 as a function of $p^{(a)}({\bm x}) \equiv {\bm e}^{{(a)}} ({\bm x})[{\bm p}]$
 [or $q({\bm x})$ and $n^{(i)}({\bm x})$].
Eventually, we will prefer to use $q$ and $n^{(i)}$ rather than $p^{(i)}$.
From the definition of $q$, Eq (\ref{def:q})
 we then obtain the desired transformation relation
 \begin{align}
 \tilde{q}\circ \tilde c{(\tilde{\bm x})}
 &\equiv q ({\bm x})+\delta q ({\bm x})
 \equiv q ({\bm x})[1+\delta \ln q ({\bm x})] \,,
 \end{align}
 as
\begin{align}\label{EqTp}
 \tilde{q}\circ \tilde c{(\tilde{\bm x})}
 &\equiv a^2(\tilde \eta) \tilde{N} p^{\tilde{0}}
 \circ \tilde c{(\tilde{\bm x})} \notag\\
 &= q ({\bm x}) \left[ 1 + \calH T + T_{,i}n^{(i)}
 + \frac{1}{2} (\calH' + \calH^2) T^2 + \calH T T_{,i}n^{(i)}
 - T' T_{,i}n^{(i)} + \frac{1}{2}  T^{,i} T_{,i}
 + T_{,i} \left( \alpha n^{(i)} - h^i{}_j n^{(j)} \right)\right] \,,
\end{align}
As for $n^{(i)}$ from Eq (\ref{def:n^i}), its transformation
\begin{align}
 n^{\widetilde{(i)}}\circ \tilde c{(\tilde{\bm x})}
 &\equiv  n^{(i)}({\bm x}) + \delta n^{(i)}({\bm x}) \,,
\end{align}
 is given by
 \begin{align}
 n^{\widetilde{(i)}}\circ \tilde c{(\tilde{\bm x})}
 &\equiv \left[ \tilde{\beta}^i + \left( \frac{1}{\tilde{N}} \delta^i{}_j
 + \tilde{h}^i{}_j \right) \frac{p^{\tilde{j}}}{p^{\tilde{0}}} \right]
 \circ \tilde c{(\tilde{\bm x})} \notag\\
 &= n^{(i)}({\bm x}) + S^{(i)(j)} T_{,j} + L^{[i,j]}  n_{(j)} \,.
 \end{align}
Here when the argument is not specified, it is $({\bm x})$.
Note that for future use, we have defined in these expressions the differences
 $\delta q$, $\delta \ln q$, and $\delta n^{(i)}$.
The first order and second order perturbation of these can be read directly
from the expressions above.
We also define $\delta q^{(i)}\equiv q\Bigl[(\delta \ln q) n^{(i)}
 + \delta n^{(i)}\Bigr]$.

It is worth stressing that these differences are measured at the same point.
For instance $\delta n^{(i)}({\bm x})=  n^{\widetilde{(i)}} \circ
 \tilde c{(\tilde{\bm x})} - n^{(i)}({\bm x}) $,
 and they do not vanish because in one case we use the tetrads
 ${\bm e}^{(i)}$ associated with the coordinate system $c$  to
 obtain the components, and in another case we use the tetrads
 $\tilde{\bm e}^{(i)}$ associated with the coordinate system $\tilde c$.
It is the basis at a given point of space-time that changes when we change
the coordinate system, not the momentum itself. This point is
illustrated in the bottom right part of Fig.~(\ref{fig1})

\subsection{Scalar distribution function}

If we were using  the natural basis associated with a coordinate system (the canonical basis) for the tangent space, then any scalar function on the tangent bundle $T{\cal M}$,
that is a function of the space-time position and of the tangent space
at each point, would transform as
\begin{equation}
I \circ \tilde c{(\tilde {\bm x},p^{\widetilde{\mu}})} = I({\bm x},p^\mu) \,,
\end{equation}
where $\tilde {\bm x}$ and ${\bm x}$ are related by~\eqref{Trulexmu}
and $p^{\widetilde{\mu}}$ and $p^{\mu}$ are related by~\eqref{Transfopmu}.
Again this rule is a statement that the function is invariant
under a change of coordinates because it is defined purely geometrically.

However, as mentioned earlier, we use the basis of the tetrad field,
${\bm e}_{(i)}$ and ${\bm e}^{(i)}$, to obtain the components of momentum
not the canonical basis. The tetrads are also completely determined by
the choice of coordinates due to our prescription~\eqref{tetrads}.
The tetrad field, though being of tensorial nature, is not invariant
as in Eq.~\eqref{GTTensor}. Furthermore, as mentioned earlier, we also work with the conformal momentum ${\bm q}$ rather than the momentum itself ${\bm p}$.
Given this choice for the basis of the tangent space, the scalar function transforms as
\begin{equation}\label{Trulef}
 I\circ \tilde c{(\tilde {\bm x},q^{\widetilde{(\imath)}})}
 = I ({\bm x},q^{(\imath)})\,,
\end{equation}
that is, it is unchanged when it is evaluated at the same point of
the tangent bundle. On the other hand, the gauge transformation is
a transformation rule at the same coordinate point and it is the relation
between $I\circ \tilde c{({\bm x},q^{{(i)}})}$ and $I{( {\bm x},q^{{(i)}})}$.
We then need the expressions of $q^{\widetilde{(\imath)}}\circ \tilde c{(\tilde{\bm x})}$
in terms of $q^{{(\imath)}}({\bm x})$ in spherical coordinates, which are derived in the previous section in~\eqref{EqTp}.
At first order, using the fact that the background distribution function
cannot depend on the direction $n^{(i)}$, we obtain
\begin{equation}
 I \circ \tilde c{({\bm x},q,n^{(i)})}
 + \left(\xi^\mu\pa_\mu +\delta q \frac{\pa }{\pa q}\right) I \circ
 \tilde c{({\bm x},q,n^{(i)})}= I ({\bm x},q,n^{(i)})\,.
\end{equation}
Given that the background distribution function also depends neither
on time nor on space but only on $q$, we obtain
\begin{equation}
 I^{(1)}\circ \tilde c{({\bm x},q,n^{(i)})}
 = I^{(1)} ({\bm x},q,n^{(i)})-\delta \ln q \,\parqof{\bar I{(q)}}\,.
\end{equation}
At second order, we obtain
\begin{equation}
 I \circ \tilde c{({\bm x},q,n^{(i)})}
 + \left(\xi^\mu\pa_\mu +\delta q \frac{\pa }{\pa q} +\delta n^{(i)} D_{(i)}
 +\frac{1}{2}\delta q^2\frac{\pa^2 }{\pa q^2}\right) I \circ
 \tilde c{({\bm x},q,n^{(i)})}= I ({\bm x},q,n^{(i)}) \,.
\end{equation}
Using the first order expressions, the gauge transformation rule reads
\begin{eqnarray}\label{GaugeTRuleI2}
 \frac{1}{2}I^{(2)}\circ \tilde c{({\bm x},q,n^{(i)})}
 &=& \frac{1}{2} I^{(2)} ({\bm x},q,n^{(i)})
 - \left(\xi^\mu\pa_\mu +\delta \ln q \parq+\delta n^{(i)} D_{(i)}\right)
 I^{(1)}{({\bm x},q,n^{(i)})}\nonumber\\*
 &&+ \left(\xi^\mu\pa_\mu +\delta n^{(i)}\frac{\pa}{\pa n^{(i)}}\right)
 (\delta \ln q) \parq \bar I{(q)}
 +\frac{1}{2}(\delta \ln q)^2 \left(\calD_q^2-2 \parq \right)\bar I(q)\,,
\end{eqnarray}
where we used $\pa \delta \ln q / \pa \ln q = 0$.

This transformation rule can be applied to $I$ or $V$
 since these are scalar valued distribution functions.
However we are interested in the transformation rule
 for the spectral components of $I$.
Using the decomposition Eq~(\ref{intensity}) for $I$ we obtain that
 the temperature is transforming under a gauge transformation as
 (noting for simplicity $\tilde \Theta \equiv \Theta \circ \tilde c$)
\begin{align}
 \tilde{\Theta}^{(1)} &= \Theta^{(1)} + (\delta \ln q)^{(1)} \,,\\
 \frac{1}{2} \tilde{\Theta}^{(2)} &= \frac{1}{2} \Theta^{(2)}
 + \frac{1}{2} (\delta \ln q)^{(2)} +\Theta^{(1)} (\delta \ln q)^{(1)}
 - \left(\xi^\mu \frac{\pa}{\pa x^\mu}+\delta n^{(i)} D_{(i)} \right)
 \Bigl[ \Theta^{(1)}+(\delta \ln q)^{(1)} \Bigr]\,,
\end{align}
 where it is implied that all quantities are evaluated either at ${\bm x}$ or
 at $({\bm x},q,n^{(i)})$.
The detailed form of the transformation rule can then be obtained just
by considering the perturbations of $q$ and $n^i$, $\delta \ln q$ and $\delta n^{(i)}$,
which have been obtained in Eqs.~(\ref{EqTp}).
For completeness we report it here
\begin{align}
 \tilde{\Theta}
 &= \Theta + \calH T + T_{,i}n^{(i)} + \frac{1}{2} (\calH^2 - \calH') T^2
 + \calH T (T_{, i} n^{(i)} - T') - (T' T_{,i} + T T_{, i}') n^{(i)} \notag\\
 &\qquad
 + T_{,i} \Bigl( \alpha n^{(i)} - h^i{}_j n^{(j)} \Bigr) - L^i (\calH T_{, i}
 + T_{, i j} n^{(j)}) + \left( n^{(i)}n^{(j)} - \frac{1}{2} \delta^{i j}
 \right) T_{,j} T_{, i} - L^{[i}{}_{,j]} T_{, i} n^{(j)} \notag\\
 &\qquad
 + (\calH T + T_{, i} n^{(i)}) \Theta - \xi^\mu \Theta_{, \mu}
 - \Bigl( S^{(i) (j)} T_{,j} + L^{[i}{}_{,j]} n^{(j)} \Bigr) D_{(i)} \Theta \,.
\end{align}
Finally, the gauge transformation of $y$ is trivial.
Since $y$ vanishes at first order, $y$ is gauge invariant at second order, and it can also be checked directly by extracting $y$ out of the transformation rule of $I$ at second order~\eqref{Trulef}.

\subsection{Baryons fluid description}

In order to obtain the complete gauge transformation of the collision
term, we need the gauge transformation rule of the baryons fluid
velocity in the tetrad frame up to second order, since it appears in
the collision term.
We obtain
\begin{align}
 v^{\widetilde{(\imath)}}\circ \tilde c{({\bm x})}
 &= v^{(i)}({\bm x}) + T^{, i} + \alpha T^{, i} - h^i{}_j T^{, j} - T' T^{, i}
 + L^{[i}{}_{, j]} (v^{(j)} + T^{, j})
 - T (v^{(i)}{}' + T^{, i}{}') - L^j (v^{(i)} + T^{, i})_{, j} \,.
\end{align}
We also need the gauge transformation rule up to first order of the
electrons density and it is easily obtained to be
\begin{equation}
 \delta_e\circ \tilde c{({\bm x})}
 = \delta_e({\bm  x}) + 3 {\cal H} T n_e \,.
\end{equation}

\subsection{Gauge dependence of the Boltzmann equation and Gauge invariant form}
\label{ssec:lc}
Having derived all the necessary gauge transformation rules,
 it is now possible to check the gauge dependence of the derived
 second order Boltzmann equation and collision term explicitly.
More precisely we shall check that the Liouville and the collision terms
 of the Boltzmann equation transform as they should do.
The Liouville term and the Collision term are distribution functions
 (scalar or tensor valued depending whether or not we are considering
 the intensity or polarisation).
From the transformation rule of
 a scalar distribution function~\eqref{GaugeTRuleI2} and
 the spectral decomposition (\ref{spectraldecI})
 we deduce that ${\cal L}^Y$ and $\calC^Y$ are gauge invariant and
 $\calL^\Theta$ and $\calC^\Theta$ should transforms up to the second order as
 (noting $\widetilde{\calL^\Theta} \equiv \calL^\Theta \circ \tilde c$ and
 $\widetilde{\calC^\Theta} \equiv \calC^\Theta \circ \tilde c$ )
 \begin{align}
 \widetilde{\calL}^\Theta
 &= \calL^\Theta - \Bigl( \xi^\mu \pa_\mu + \delta n^{(i)} D_{(i)} \Bigr) \calL^{\Theta\,(1)}
 + 2 \calH T \calL^{\Theta\,(1)} \,, \\
 \widetilde{\calC^\Theta}
 &= \calC^\Theta - \Bigl( \xi^\mu \pa_\mu + \delta n^{(i)} D_{(i)} \Bigr) \calC^{\Theta (1)}
 + 2 \calH T \calC^{\Theta (1)} \,.
 \end{align}

After very long and tedious but straightforward calculations using all the transformation rules derived so far for the distribution function, the metric components and
the baryons velocity and energy density, we have checked that the Liouville operator
$\calL^\Theta$ and the collision term $\calC^\Theta$ actually transform as in the above equations.
This completes the consistency test of the Boltzmann equation that we
have derived as well as all the gauge transformation rules obtained
for its constituents.

The same property is found of course for the circular polarisation
since in that case the collision term vanishes.
As for polarisation, we have also checked that the Liouville operator
 $\calL^P_{(i)(j)}$ and the collision term $\calC^P_{(i)(j)}$
transform like tensor valued quantities (see the details in Appendix~\ref{AppGTLC}). More importantly, we have
checked that the Liouville and collision terms for the spectral distortions ($\calL^Y_{(i)(j)}$ and $\calC^Y_{(i)(j)}$) are gauge invariant as it should be since they vanish on the background and first-order spacetimes.

The gauge invariance of the Boltzmann equation, as in the case of Einstein equation and in general for covariant equations, enables us to write it down in terms of gauge invariant variables. In practice, it is equivalent to completely fix the gauge and write down
the equations in term of the perturbation in this gauge.

\section{Summary and discussion}
\label{sec:conc}

In this paper we derived the second order Boltzmann equation
in the most general manner incorporating polarisation and
without fixing a gauge. In order to describe the polarisation of photon, we used a formalism
based on a tensor-valued distribution function. We performed the
separation between temperature and spectral distortion for the
intensity and we also extended this separation to polarisation.

We then derived the gauge transformation rules for the metric, the
momentum and the distribution function to see how those quantities are
mixed under the gauge transformation. As an application, we checked
the gauge dependence of the derived Boltzmann equation under a gauge
transformation and obtained consistent transformation rules. This is a
non-trivial check of the correctness of the derived equations as well
as the gauge transformation rules.

We now discuss two issues related to the gauge dependence in the Boltzmann equation.

\subsection{Gauge dependence of lensing term}

It is well known that the lensing term
\begin{align}
 \calL \supset \frac{\dd n^{(i)}}{\dd \lambda} D_i \Theta
 ~ : ~ (\ma{lensing ~ term}) \,,
\end{align}
which is written in terms of the conventional lensing potential
in the Newtonian gauge significantly affects the bispectrum of CMB~\cite{Hanson:2009kg,Lewis:2011fk,Ade:2013ydc}.
Indeed, the correlation between the lensing and ISW effect is the
dominant contribution to the bias for the local-type non-Gaussianity
in Planck. However it is very hard to include this contribution in the
line of sight integration and evaluate it until today. Usually, the
lensing effect is added separately to the final result obtained in the
Poisson gauge. However the effect from this lensing term depends on
the gauge choice. Actually, as we have seen above, the lensing term is
mixed with other terms under the gauge transformation. This means that
some of lensing effects in a specific gauge are absorbed into other
effects in another gauge. In principle, there exists a gauge where we
can avoid the difficult computation of this lensing term to some
extent by evaluating other more tractable terms. Since we have derived
the Boltzmann equation without choosing any specific gauge, it should be possible to investigate this possibility further.

\subsection{Observed temperature anisotropies}

Here we make a comment on the observed temperature anisotropies.
In the main part of this paper, we have shown that the second order Boltzmann equation
is gauge invariant and thus it can be written in terms of gauge invariant quantities.
However, there is a subtlety in the meaning of "gauge invariance".
This originates from our choice of the local inertial frame.

As is clear from Eq~(\ref{tetrads}), we always choose the local inertial frame so that the three-velocity vanishes, $\hat{v}^i=0$ and there is no rotation of the spatial axis relative to the background spatial coordinate axis, let us call $\theta_i=0$.
In order to achieve this, the local inertial frame
has to be changed when we perform a gauge transformation.
If we were to identify this local inertial frame as the one of an observer,
we would be lead to consider different observers in different gauges.
This is clear from the gauge transformation at the first order:
\begin{equation}
 \Theta \to \Theta + {\cal H} T + T_{,i} n^i\,.
\label{1sttempg}
\end{equation}
The last term comes from a change of local inertial frame.
By fixing the gauge we can promote $\Theta$ to the gauge invariant temperature fluctuations
but these are temperature fluctuations observed by an observer
with $\hat{v}^i=0$ and $\theta_i=0$ in this gauge.
In order to evaluate temperature fluctuations observed by a different
observer, we need to change the local inertial frame. Alternatively,
we can perform a gauge transformation keeping the conditions $\hat{v}^i=0$ and $\theta_i=0$ for the local inertial frame
 so that this frame coincides with the one of the observer.

In all the literature, the second order temperature anisotropies are
calculated in the Poisson gauge so far, with a specific choice of
the local inertial frame. Strictly speaking this is not the temperature anisotropies that
we observe as there is no reason for us to be comoving
with the local inertial frame associated with such a gauge.
One thus needs to change the local inertial frame or change a gauge.
At first order, this was not an issue. As is clear from~(\ref{1sttempg}),
the change of gauge and
the local inertial frame only affect the monopole $\ell =0$ and
dipole $\ell =1$ if we expand the temperature anisotropies into
multipole components. Thus, the $\ell \geq 2$ modes are not affected by the change of observers. However this is no longer the case at the second order.
In the second order gauge transformation, there are terms that
are convolutions of the first order temperature anisotropies and
the gauge transformation;
 \begin{align}
 \Theta \to \Theta - \xi^\mu \Theta_{, \mu} - \delta n^{(i)} D_{(i)} \Theta
 + \cdots \,.
 \end{align}
These terms affect the observed temperatures even for the $\ell \geq 2$ modes.

In order to define the "observed temperature anisotropies'', we should
keep the conditions $\hat{v}^i=0$ and $\theta_i=0$ for the local
inertial frame and specify the gauge so that this local inertial frame coincides with our local inertial frame
where we perform experiments. Thus special care must be taken when we compare theoretical predictions to observations. Our formula for the gauge transformation will be useful to investigate this issue further.


\acknowledgments

A.N is grateful to Shuichiro Yokoyama and Ryo Saito
 for their continuous encouragement and fruitful discussion.
This work was supported in part by Monbukagaku-sho
Grant-in-Aid for the Global COE programs,
"The Next Generation of Physics, Spun from Universality
and Emergence'' at Kyoto University,
by JSPS Grant-in-Aid for Scientific Research (A) No.~21244033,
and by the Long-term Workshop at Yukawa Institute
on Gravity and Cosmology 2012, YITP-T-12-03.
AN is partly supported by Grant-in-Aid for JSPS Fellows No.~21-1899 and
 JSPS Postdoctoral Fellowships for Research Abroad.
CP was supported by the STFC (UK) grant ST/H002774/1
 during the first part of this research, and was then supported by French state funds managed by the ANR within the Investissements d'Avenir programme under reference ANR-11-IDEX-0004-02.
K.K. is supported by STFC grant ST/H002774/1, ST/K0090X/1,
 the European Research Council and the Leverhulme trust.

\appendix

\section{ADM variables and usual perturbation variables}
\label{app:ADM}

From the form of the metric in the ADM parametrisation~(\ref{MetricADM}),
and the perturbation of the lapse function and the spatial metric given in the equations~(\ref{DefPertNNi}),
the perturbed metric is expressed up to second order as
 \begin{equation}
 \dd s^2 = a^2 (\eta) \Bigl[ - (1 + 2 \alpha + \alpha^2 - \beta_i \beta^i)
 \dd \eta^2 + 2 (\delta_{i j} + 2 h_{i j}) \beta^j \dd x^i \dd \eta
 + (\delta_{i j} + 2 h_{i j}) \dd x^i \dd x^j \Bigr] \,.
 \end{equation}
This has to be compared with the usual parametrisation of the
perturbations of the metric which is in the form
 \begin{gather}
 \dd s^2 = a^2 (\eta) \Bigl[ - (1 + 2 A) \dd \eta^2 + 2 B_i \dd x^i \dd \eta
 + (\delta_{i j} + 2 C_{i j}) \dd x^i \dd x^j \Bigr] \,,
 \end{gather}
where $C_{i j}$ can be further split into 2 scalar, 2 vector and 2 tensor
degrees of freedom. By a direct comparison of these two parametrisations,
the relation between the two metric parametrisations is
\begin{subequations}
\begin{align}
 g_{0 0} &: 1 + 2 \alpha + \alpha^2 - \beta_i \beta^i = 1 + 2 A\,,\\
 g_{0 i} &: \beta_i + 2 h_{i j} \beta^j = B_i\,, \\
 g_{i j} &: \delta_{i j} + 2 h_{i j} = \delta_{i j} + 2 C_{i j}\,.
\end{align}
\end{subequations}
At first order we obtain
\begin{gather}
 A^{(1)} = \alpha^{(1)} \,, \qquad B_i^{(1)} = \beta^{(1)}_i \,, \qquad
 C_{i j}^{(1)} = h_{i j}^{(1)} \,,
\end{gather}
and the two parametrisations are the same. However, at second order we get the relations
\begin{gather}
 A^{(2)}=  \alpha^{(2)} + \alpha^{(1) 2} - \beta^{(1)}_i \beta^{(1) i}  \,, \qquad
 B_i^{(2)} = \beta^{(2)}_i + 4 h_{i j}^{(1)} \beta^{(1) j}  \,,\qquad
 C_{i j}^{(2)} = h_{i j}^{(2)} \,.
\end{gather}

\section{Construction of the distribution function for polarised light}
\label{app:Dist}

We consider a two-dimensional polarisation plane defined by two unit complex vectors $\hat \bepsilon_{(\ma{I})}$ and $\hat \bepsilon_{(\ma{II})}$, which are mutually orthogonal,
 $\hat \epsilon_{(\ma{A})}{}^{\star\mu} \hat \epsilon_{(\ma{B})}{}^\nu g_{\mu\nu}
 =\delta_{(\ma{A})(\ma{B})}$. Any polarisation $\boldsymbol{\epsilon}$ can be represented
by a superposition of $\hat \bepsilon_{(A)}$ in the form
\begin{equation}
  \bepsilon = \sum_{A=\ma{I},\ma{II}} \epsilon^A \hat \bepsilon_{(A)}
 \label{pol}\,.
 \end{equation}
The orthogonal vectors $\hat \bepsilon_{(A)}$ define a polarisation plane and
we choose them to be orthogonal to the direction of the photon $n^{(i)}$ and
to the observer velocity ${\bm e}_{(0)}$,
\begin{equation}\label{EqProjepsilons}
 \hat \epsilon_{(A)}{}^\mu n^{\nu} g_{\mu \nu}
 = \hat \epsilon_{(A)}{}^\mu {e}_{(0)}{}^\nu g_{\mu\nu}=0\,.
\end{equation}
We can also associate canonically polarisation forms through $\hat \epsilon^{(A)}{}_\mu \equiv g_{\mu\nu} \hat \epsilon_{(A)}{}^\nu$, and they will be also complex unit forms and mutually orthogonal. The polarisation density matrix $f_{A B}$ is defined so that the expected number of photon in a phase-space element with a polarisation state $\bepsilon$ is given by
 \begin{gather}
 f( {\bf x}, {\bf p}, \bepsilon)
 \equiv f_{A B}({\bf x}, {\bf p}) \epsilon^{\star A} \epsilon^B \,.
 \end{gather}
With such a parametrisation, all the electromagnetic gauge degrees of freedom
have been fixed and we parametrise the physical degrees of freedom of this density matrix by the usual Stokes parameters as
\begin{gather}
 f_{A B} = \frac{1}{2}
\begin{pmatrix}
 I + Q & U - i V \\
 U + i V & I - Q
\end{pmatrix} \,,
\end{gather}
where it is implied that $f_{AB}$ and the Stokes parameters depend on
the position $x^\mu$ and on the momentum $p^\mu$ [or $(q,n^{(i)})$ in spherical coordinates]. $f_{AB}$ is a Hermitian matrix since $f_{AB}=f_{BA}^\star$.

From the four-dimensional point of view, the polarisation density matrix is a tensor-valued distribution function. It is a 2-form defined by
\begin{gather}
 f_{\mu \nu} \equiv f_{A B} \epsilon^{\star(A)}{}_\mu \epsilon^{(B)}{}_\nu \,,
\end{gather}
and the expected number of photon in a phase-space element
for a polarisation state $\bepsilon$ is given by
\begin{equation}\label{Eqdefffromfmunu}
 f ({\bf x}, {\bf p}, \bepsilon)
 \equiv f_{\mu\nu} ({\bf x}, {\bf p}) \epsilon^{\star\mu} \epsilon^\nu \,.
\end{equation}
This can be viewed as a multipolar expansion in the polarisation state.
From~(\ref{EqProjepsilons}), the tensor-valued distribution function is
a projected quantity such that
\begin{equation}
f_{\mu \nu} =S_\mu{}^\alpha S_{\mu}{}^\beta f_{\alpha \beta}\,.
\end{equation}
It is then straightforward to realize that it can be decomposed according to~(\ref{Pol}).

\section{Boltzmann equation for the tensor-valued distribution function}
\label{app:Boltz_pol}

\subsection{From a scalar valued to a tensor-valued distribution function}

In this section, we explain in detail how the Boltzmann equation for the tensor-valued
distribution function can be obtained from the Boltzmann equation of a scalar distribution function. Since this scalar distribution function $f$ depends on $x^\mu,p^\mu$
but also on $\epsilon^\mu$, the action of the Liouville operator is given by
\begin{equation}\label{CrudeBoltzmann}
 \frac{\calD}{\calD \lambda} f (x^\mu, p^\mu, \epsilon^\mu)
 = \frac{\dd x^\alpha}{\dd \lambda} \frac{\pa f}{\pa x^\alpha}
 + \frac{\dd p^\alpha}{\dd \lambda} \frac{\pa f}{\pa p^\alpha}
 + \frac{\dd \epsilon^\alpha}{\dd \lambda} \frac{\pa f}{\pa \epsilon^\alpha}
 = C [f]\,.
\end{equation}
In the geometric optics approximation, $p^\mu$ and $\epsilon^\mu$ are
parallel transported and we obtain
\begin{align}
 0 =& \frac{\calD p^\alpha}{\calD \lambda}
 = \frac{\dd p^\alpha}{\dd \lambda}
 + \Gamma^\alpha_{\beta \gamma} p^\beta p^\gamma \,, \\
 0 =& \frac{\calD \epsilon^\alpha}{\calD \lambda}
 = \frac{\dd \epsilon^\alpha}{\dd \lambda}
 + \Gamma^\alpha_{\beta \gamma} \epsilon^\beta p^\gamma \,.
\end{align}
Using~(\ref{Eqdefffromfmunu}), the term involving
the evolution of polarisation is obtained as
\begin{gather}
 \frac{\dd \epsilon^\alpha}{\dd \lambda} \frac{\pa f}{\pa \epsilon^\alpha}
 = f_{\alpha \beta} \frac{\dd \epsilon^{\alpha}}{\dd \lambda} \epsilon^{* \beta}
 + f_{\alpha \beta} \frac{\dd \epsilon^{\star \beta}}{\dd \lambda} \epsilon^{\alpha}
 = - \Gamma^{\alpha}_{\gamma \delta} \epsilon^\gamma p^\delta f_{\alpha \beta}
 \epsilon^{* \beta}
 - \Gamma^{\beta}_{\gamma \delta} \epsilon^{\star\gamma} p^\delta f_{\alpha \beta}
 \epsilon^{\alpha}\,.
\end{gather}
Combining this result with the space-time derivative term of the Liouville operator, we get
\begin{gather}
 p^\alpha \frac{\pa f}{\pa x^\alpha} + \frac{\dd \epsilon^\alpha}{\dd \lambda}
 \frac{\pa f}{\pa \epsilon^\alpha}
 = p^\gamma \epsilon^\alpha (\nabla_\gamma f_{\alpha \beta}) \epsilon^{* \beta} \,,
\end{gather}
and thus the Boltzmann equation~(\ref{CrudeBoltzmann}) can be
rewritten as
\begin{equation}\label{EqBoltzmannfee}
 \frac{\calD f}{\calD \lambda}
 = \epsilon^\mu \left( p^\alpha \nabla_\alpha f_{\mu \nu}
 + \frac{\dd p^\alpha}{\dd \lambda} \frac{\pa f_{\mu \nu}}{\pa p^\alpha} \right)
 \epsilon^{* \nu}
 = C [f] \equiv \epsilon^\mu C_{\mu \nu} \epsilon^{* \nu} \,.
\end{equation}
The last equality defines the tensor-valued collision term.
If we do fix the electromagnetic gauge condition for the collision term
in the same manner as what we did for $f_{\mu \nu}$,
that is, if $C_{\mu \nu} =S_\mu{}^\alpha S_{\nu}{}^\beta C_{\alpha \beta}$,
then the Boltzmann equation for $f_{\mu\nu}$ is given by
\begin{equation}\label{EqBaseBoltzmann}
 S_\mu{}^\alpha S_{\mu}{}^\beta \frac{\calD f_{\alpha \beta}}{\calD \lambda}
 \equiv S_\mu{}^\alpha S_{\mu}{}^\beta \left(p^\sigma \nabla_\sigma f_{\alpha \beta}
 + \frac{\dd p^\sigma}{\dd \lambda}
 \frac{\pa f_{\alpha \beta}}{\pa p^\sigma}\right)
 = C_{\mu\nu}\,.
\end{equation}
Note that the use of the projectors is required because the components of the equation
which are not in the polarisation plane are not fixed by~(\ref{EqBoltzmannfee}).

\subsection{From the canonical basis to the tetrad basis}
\label{sapp:covp}

In the equation~(\ref{EqBaseBoltzmann}), the Greek indices refer to a
given coordinate system and its canonical basis for the tangent
space, and the distribution function is a function of $(x^\mu,p^\mu)$.
If we want to use instead an orthonormal basis for the tangent space,
that is, to use the components $p^{(i)}=e^{(i)}{}_\mu p^\mu$ or
the conformal momentum components $q^{(i)} = a p^{(i)}$ in the tetrad basis,
then the Boltzmann equation can be modified accordingly.
In order to do so, we need to be explicit about the partial derivatives to emphasize which
variables are to be kept constant when the partial derivatives are evaluated.
The Boltzmann equation reads indeed
\begin{equation}\label{Boltzbaseopen}
S_\mu{}^\alpha S_{\nu}{}^\beta \frac{\calD f_{\alpha \beta}}{\calD \lambda}
 = S_\mu{}^\alpha S_{\nu}{}^\beta \left( p^\gamma
 \left.\frac{\pa f_{\alpha \beta}}{\pa x^\gamma}\right|_{p^\mu}
 - p^\gamma \Gamma^\delta_{\gamma \alpha} f_{\delta \beta}
 - p^\gamma \Gamma^\delta_{\gamma \beta} f_{\alpha \delta}
 + \left.\frac{\dd p^\gamma}{\dd \lambda}
 \frac{\pa f_{\alpha \beta}}{\pa p^\gamma}\right|_{x^\mu}\right) \,.
\end{equation}
Using the properties
\begin{equation}
 \left.\frac{\pa f_{\alpha \beta}}{\pa x^\mu} \right|_{p^\mu}
 = \left.\frac{\pa f_{\alpha \beta}}{\pa x^\mu}\right|_{q^{(i)}}
 + \frac{\pa  f_{\alpha \beta}}{\pa q^{(i)}}
 \frac{\pa (a e^{(i)}{}_\nu)}{\pa x^\mu}p^\nu \,, \qquad
 \left.\frac{\pa f_{\alpha \beta}}{\pa p^\mu}\right|_{x^\mu}
 = \left.\frac{\pa f_{\alpha \beta}}{\pa q^{(i)}}\right|_{x^\mu} ae^{(i)}{}_\mu \,,
 \qquad
 \frac{\dd q^{(i)}}{\dd \lambda}
 = a e^{(i)}{}_\mu \frac{\dd p^\mu}{\dd \lambda}
 + p^\mu \frac{\dd (a e^{(i)}{}_\mu)}{\dd \lambda} \,,
\end{equation}
we then deduce that
\begin{align}\label{Boltzbasetetrad}
 S_\mu{}^\alpha S_{\nu}{}^\beta \frac{\calD f_{\alpha \beta}}{\calD \lambda}
 &=S_\mu{}^\alpha S_\nu{}^\beta  \left(p^\gamma
 \left.\frac{\pa f_{\alpha \beta}}{\pa x^\gamma} \right|_{q^{(i)}}
 - p^\gamma \Gamma^\delta_{\gamma \alpha} f_{\delta \beta}
 - p^\gamma \Gamma^\delta_{\gamma \beta} f_{\alpha\delta}
 + \left.\frac{\dd q^{(i)}}{\dd \lambda}
 \frac{\pa f_{\alpha \beta}}{\pa q^{(i)}} \right|_{x^\mu}\right) \notag\\
 &= S_\mu{}^\alpha S_{\mu}{}^\beta
 \left(p^\gamma \nabla_\gamma f_{\alpha \beta} + \frac{\dd q^{(i)}}{\dd \lambda}
  \frac{\pa f_{\alpha \beta}}{\pa q^{(i)}}\right)\,.
\end{align}
Comparing Eqs.~(\ref{Boltzbaseopen}) and (\ref{Boltzbasetetrad}),
we notice that the notation $p^\gamma \nabla_\gamma f_{\alpha \beta}$
 could be ambiguous.
Indeed if we use the canonical coordinate system for the tangent space,
then $\pa/\pa x^\mu$ is to be taken at $p^\mu$ fixed, but if we take
the tetrad basis (or another coordinate system for the tangent space),
then $\pa/\pa x^\mu$ is to be taken with $q^{(i)}$ fixed.

Eq.~(\ref{Boltzbasetetrad}) is not exactly the desired form of the Boltzmann
equation when the distribution function depends on $(x^\mu,q^{(i)})$.
In fact, the use of the tetrad basis makes it natural to work
with spherical coordinates in the tangent space.
In order to introduce them, we first relate the Cartesian derivative
in the tangent space, that is, the derivative with respect to $q^{(i)}$,
to the covariant derivative on the unit sphere which is described
by the possible directions $n^{(i)}$ of the momentum.
We must stress that at any point of space-time, $x^\mu$,
a distribution function (tensor-valued like $f_{\mu\nu}$ or
scalar valued like its trace $I$) which depends on $(x^\mu,q^{(i)})$
can be considered as a field in the tangent space because the tangent space
at a given point can be considered as a flat three-dimensional manifold whose points are labelled by $q^{(i)}$ and the natural covariant derivative in this manifold is
$\pa/\pa q^{(i)}$.
Using that $q^{(i)} = q n^{(i)}$, and the property
\begin{equation}\label{ExtrinsicK}
\frac{\pa n^{(i)}}{\pa q^{(j)}} = \frac{1}{q} S^{(i)}{}_{(j)} \,,
\end{equation}
it is possible to show the following relations;
\begin{align}\label{Expandderpi}
 q \frac{\pa T_{\mu\nu}}{\pa q^{(i)}}
 &= \frac{\pa T_{\mu\nu}}{\pa \ln q} n_{(i)} + D_{(i)} T_{\mu\nu}
 - e_{(i)}{}^\rho (T_{\mu \rho} n_{\nu} + T_{\nu \rho}n_{\mu}) \,,
 &\quad\text{with}\quad
 & D_{(i)} T_{\mu\nu} \equiv q S_{(i)}{}^{(j)} S_\mu{}^\alpha S_\nu{}^\beta
 \frac{\pa T_{\alpha\beta}}{\pa q^{(j)}} \,, \\*
 q \frac{\pa S}{\pa q^{(i)}}
 &= \frac{\pa S}{\pa \ln q} n^{(i)} + D_{(i)} S \,,
 &\quad\text{with}\quad
 & D_{(i)} S \equiv q S_{(i)}{}^{(j)} \frac{\pa S}{\pa q^{(j)}} \,,
\end{align}
where $T_{\mu\nu}(q^{(i)})$ is a projected tensor field in the tangent space
such that $T_{\mu\nu}(q^{(i)}) n^{\nu}=T_{\mu\nu}(q^{(i)}) n^{\nu}=0$,
and $S(q^{(i)})$ is a scalar field in the tangent space.
Here $D_{(i)}$ is the covariant derivative on the two-sphere
associated with the unit direction vector $n^{(i)}$,
and it appears naturally as an induced derivative on the sphere,
given that this is the surface orthogonal to $n^{(i)}$
(see Ref.~\cite{Gourgoulhon:2007ue} for more details on induced derivatives).
We can then deduce the useful property
\begin{equation}\label{KeyRuleForCovDSpherical}
 q S_\mu{}^\alpha S_\nu{}^\beta \frac{\pa f_{\alpha \beta}}{\pa q^{(i)}}
 = \frac{\pa f_{\mu\nu}}{\pa \ln q} n_{(i)} + D_{(i)} f_{\mu\nu} \,.
\end{equation}
Given that
\begin{equation}
\frac{\dd q^{(i)}}{\dd \lambda}
 = q \left(\frac{\dd \ln q }{\dd \lambda} n^{(i)}
 + \frac{\dd n^{(i)}}{\dd \lambda} \right)\,,
\end{equation}
from Eqs.~(\ref{Expandderpi}) and Eq.~(\ref{Boltzbasetetrad}),
we then find that the Boltzmann equation takes the form
\begin{equation}
 S_\mu{}^\alpha S_\nu{}^\beta \frac{\calD f_{\alpha \beta}}{\calD \lambda}
 = S_\mu{}^\alpha S_\nu{}^\beta \nabla_\gamma f_{\alpha \beta}
 \frac{\dd x^\gamma}{\dd \lambda}
 + \frac{\pa f_{\mu \nu}}{\pa \ln q} \frac{\dd \ln q}{\dd \lambda}
 + D_{(i)} f_{\mu \nu} \frac{\dd n^{(i)}}{\dd \lambda} = C_{\mu \nu} \,.
\end{equation}

\subsection{Decomposition of the Boltzmann equation}

In order to obtain equations for the components of $f_{\mu\nu}$, $I, P_{\mu \nu}$ and  $V$, we want to apply the same type of decomposition on the equation itself. Applying $\calD/\calD \lambda$ on the decomposition~\eqref{Pol} of
$f_{\mu\nu}$ leads to
\begin{gather}
 \frac{\calD f_{\mu \nu}}{\calD \lambda}
 = \frac{1}{2} \left( \frac{\calD I}{\calD \lambda} S_{\mu \nu}
 + I \frac{\calD S_{\mu \nu}}{\calD \lambda} \right)
 + \frac{\calD P_{\mu \nu}}{\calD \lambda}
 + \frac{\ii}{2} \epsilon_{\alpha \mu \nu \beta}
 \left( \frac{\calD V}{\calD \lambda} e_{(0)}{}^\alpha n^\beta
 + V \frac{\calD (e_{(0)}{}^\alpha n^\beta)}{\calD \lambda} \right) \,.
\end{gather}
We then need to screen-project this equation in order to obtain
the tensor-valued Boltzmann equation. Then the last two terms vanish.
Indeed, first
\begin{equation}
 S_\mu{}^\alpha S_\nu{}^\beta \frac{\calD S_{\alpha \beta}}{\calD \lambda}
 = S_\mu{}^\alpha S_\nu{}^\beta \left[ \frac{\dd p}{\dd \lambda} \left(
 - \frac{p_\alpha e^{(0)}{}_\beta + e^{(0)}{}_\alpha p_\beta}{p^2}
 + 2 \frac{ p_\alpha p_\beta}{p^3} \right)
 + \frac{1}{p} \left( p_\alpha \frac{\calD e^{(0)}{}_\beta}{\calD \lambda}
 + \frac{\calD e^{(0)}{}_\alpha}{\calD \lambda} p_\beta \right) \right] = 0\,.
\end{equation}
Second, from the normalization condition of $e_{(0)}{}^\mu$ and $n^\mu$,
we can show that
\begin{equation}
 e^{(0)}{}_\mu \frac{\calD e_{(0)}{}^\mu}{\calD \lambda} = 0 \,, \qquad
 n_\mu \frac{\calD n^\mu}{\calD \lambda} = 0 \,.
\end{equation}
This means that the derivative of $e_{(0)}{}^\mu$ is orthogonal to $e^{(0)}{}_\mu$
and the derivative of $n^\mu$ is orthogonal to $n_\mu$.
We thus find that
\begin{equation}
 S_\mu{}^\alpha S_\nu{}^\beta \epsilon_{\gamma \alpha \beta \delta}
 \frac{\calD (e_{(0)}{}^\gamma n^\delta)}{\calD \lambda}
 = S_\mu{}^\alpha S_\nu{}^\beta \epsilon_{\gamma \alpha \beta \delta}
 \left( \frac{\calD e_{(0)}{}^\gamma}{\calD \lambda} n^\delta
 + e_{(0)}{}^{\gamma} \frac{\calD n^\delta}{\calD \lambda} \right) = 0 \,.
\end{equation}
Finally, we obtain that the Boltzmann equation for the tensor-valued
distribution functions can be split into the desired form as
\begin{gather}
 S_\mu{}^\alpha S_\nu{}^\beta \frac{\calD f_{\alpha \beta}}{\calD \lambda}
 = \frac{1}{2} \frac{\calD I}{\calD \lambda} S_{\mu \nu}
 + S_\mu{}^\alpha S_\nu{}^\beta \frac{\calD P_{\alpha \beta}}{\calD \lambda}
 + \frac{\ii}{2} \frac{\calD V}{\calD \lambda}
 \epsilon_{\alpha \mu \nu \beta} e_{(0)}{}^\alpha n^\beta = C_{\mu \nu} \,.
\end{gather}

\subsection{Expression of the Boltzmann equation for polarisation}
\label{sapp:Boltzeq_pol}

First of all, the Boltzmann equation for the circular polarisation is
the same as that for the intensity, but with $\bar V=0$ and with a vanishing collision term
since it is not generated by the Compton scattering.
We shall not study further the equation dictating the evolution of $V$
since it should remain null at all time unless generated by other types of collisions.

As for the linear polarisation, let us write down the basic equations
for the tetrad components as commonly done at first order.
By moving to the tetrad components, one can rewrite the covariant derivative in the Liouville operator as
\begin{align}
 e_{(a)}{}^\mu e_{(b)}{}^\nu L [{\bf P}\,]_{\mu \nu}
 &= S_{(a)}{}^{(c)} S_{(b)}{}^{(d)} \left( P_{(c) (d) | (e)}
 e^{(e)}{}_\mu \frac{\dd x^\mu}{\dd \lambda}
 + \frac{\pa P_{(c) (d)}}{\pa \ln q} \frac{\dd \ln q}{\dd \lambda}
 + D_{(i)} P_{(c) (d)} \frac{\dd n^{(i)}}{\dd \lambda} \right) \,,
\end{align}
 where
\begin{align}
 f_{(a) (b) | (c)}
 &\equiv e_{(c)}{}^\mu \pa_\mu f_{(a) (b)}
 - w^{(d)}{}_{(a) (c)} f_{(d) (b)} - w^{(d)}{}_{(b) (c)} f_{(a) (d)} \,,
\end{align}
 and $w^{(a)}{}_{(b) (c)}$ is the Ricci rotation coefficient defined by
 $w^{(a)}{}_{(b) (c)} \equiv e^{(a)}{}_\mu \nabla_{(c)}
 e_{(b)}{}^\mu$.
Here we also notice that only the spatial component has non-vanishing term
because the projection of $S_\mu{}^\nu$ onto $e^\mu{}_{(0)}$ vanishes
by construction. After the decomposition of the Boltzmann equation for polarisation,
one obtains the following equation for the temperature part as in Eq. (\ref{eq:Boltz_pol_dis}) up to the second order
\begin{align}
 \calL^P_{(i) (j)}
 &= \calP_{(i) (j)}{}' + \calP_{(i) (j), k} n^{(k)}
 + \frac{a^2}{q}\left[\left. \frac{\dd \eta}{\dd \lambda} \right|^{(1)} \calP_{(i) (j)}{}'
 + \left. \frac{\dd x^k}{\dd \lambda} \right|^{(1)} \calP_{(i) (j), k}
 + \left. \frac{\dd n^{(k)}}{\dd \lambda} \right|^{(1)} D_{(k)} \calP_{(i) (j)}\right]
 \notag\\
 & \qquad
 - a\Bigl( w^{(k)}{}_{(l) (0)} + w^{(k)}{}_{(l) (m)} n^{(m)} \Bigr)^{(1)}
 \Bigl( S_{(i)}{}^{(l)} \calP_{(k) (j)}
 + S_{(j)}{}^{(l)} \calP_{(k) (i)} \Bigr)\,.
\end{align}
As for the spectral distortion, the equation is given by
\begin{align}
 \calL^Y_{(i) (j)}
 &= Y_{(i) (j)}{}' + Y_{(i) (j), k} n^{(k)} \,.
\end{align}
The complete expression of the collision term
for the linear polarisation is given by
\begin{align}
 \calC_{(i) (j)}^P
 &= a \, \bar{n}_e \sigma_T \left(
 - \calP_{(i) (j)} - \frac{3}{4} \calT_{(i) (j)}{}^{(k) (l)}
 \Bigl[ \langle \Theta m_{(k) (l)} \rangle
 - 2 \langle \calP_{(k) (l)} \rangle \Bigr]
 + v^{(k)} n_{(k)} \calP_{(i) (j)} \right) \notag\\
 &\qquad
  + a \, \bar{n}_e \sigma_T \biggl(
 \calT_{(i) (j)}{}^{(k) (l)} \Bigl[ \calQ_{(k) (l)}^T
 - \Theta \calQ_{(k) (l)}^{T ~~~ (1)} \Bigr] 
 - 3 \calC^\Theta \calP_{(i) (j)} + \delta_e \calC^P_{(i) (j)} \biggr) \,, \\
 C^Y_{(i) (j)}
 &= a \, \bar{n}_e \sigma_T \left( - Y_{(i) (j)}
 - \frac{3}{4} \calT_{(i) (j)}{}^{(k) (l)}
 \Bigl[ \langle y m_{(k) (l)} \rangle
 - 2 \langle Y_{(k) (l)} \rangle \Bigr]
 + \calT_{(i) (j)}{}^{(k) (l)} \calQ_{(k) (l)}^Y 
 - \calC^\Theta \calP_{(i) (j)} \right) \,,
\end{align}
where $\calT_{(i) (j)}{}^{(k) (l)}$ is a traceless projection operator
with respect to $S_{(i) (j)}$
\begin{align}
 \calT_{(i) (j)}{}^{(k) (l)}
 \equiv S_{(i)}{}^{(k)} S_{(j)}{}^{(l)} - \frac{1}{2} S^{(k) (l)} S_{(i) (j)} \,.
\end{align}

Finally before closing this section let us explicitly write down the
equation for the temperature part for the sake of completeness. It is
of the form
 \begin{align}
 \calL^P_{(i) (j)} = \calC_{(i) (j)}^P \,,
 \end{align}
and using that at first order
\begin{equation}
w^{(k)}{}_{(l) (0)} = \frac{1}{a}\beta^{[k}{}_{, l]} \,,\qquad
 w^{(k)}{}_{(l) (m)} =\frac{2}{a}  h_m{}^{[k}{}_{, l]} \,,
\end{equation}
the explicit form of $\calL^P_{(i) (j)}$ is
\begin{align}
 \calL_{(i) (j)}^P \equiv
 &\calP_{(i) (j)}{}' + \calP_{(i) (j), k} n^{(k)} \,,
 - \alpha \calP_{(i) (j)}{}' - (\beta^k + h^k{}_l n^{(l)}) \calP_{(i) (j), k} 
 \notag\\
 & \qquad
 - \Bigl( \alpha_{, k} - \beta_{l, k} n^{(l)} + h_{k l}{}' n^{(l)}
 + 2 h_{m [k, l]} n^{(l)} n^{(m)} \Bigr) D^{(k)} \calP_{(i) (j)}
 \notag\\
 & \qquad \qquad
 - \Bigl( \beta^{[k}{}_{, l]} + 2 h_m{}^{[k}{}_{, l]} n^{(m)} \Bigr)
 \Bigl( S_{(i)}{}^{(l)} \calP_{(k) (j)} + S_{(j)}{}^{(l)} \calP_{(k) (i)} \Bigr) \,.
\end{align}

\section{Technical details about linear polarisation}

\subsection{Gauge transformation of a tensor-valued distribution function}
\label{AppGTtensor}

In this section, we investigate the transformation property of the
distribution matrix up to the second order. The transformation properties
of $I$ and $V$, which are scalar distribution functions, have already been
investigated in the main text. In this section we focus on the case of
the linear polarisation and set $I=V=0$ such that $f_{\mu\nu}=P_{\mu\nu}$.
Since the polarisation matrix is at least first order,
this means that we need only to keep terms which are first order in the coordinates transformation $(T,L^i)$.

First we must understand the transformation properties of the screen
projector. As for the tetrads, it is not a purely geometric quantity
and it is not invariant under a change of coordinates.
Indeed, it is defined with respect to the time-like tetrad
which depends on the choice of coordinates.
We thus have the two related screen projectors
\begin{equation}
{\bm S}({\bm p})={\bm g}+{\bm e}^{(0)} \otimes {\bm e}^{(0)}-{\bm
  n}\otimes{\bm n}\,,\quad \tilde{\bm S}({\bm p})={\bm g}+\tilde {\bm
  e}^{(0)} \otimes \tilde {\bm e}^{(0)}-\tilde{\bm
  n}\otimes\tilde{\bm n}\,,\quad\text{with}\quad {\bm n}\equiv
\frac{{\bm p}}{{p}^\mu{e}^{(0)}_\mu}-{\bm e}^{(0)}\,,\quad \tilde{\bm n}\equiv
\frac{{\bm p}}{{p}^\mu \tilde {e}^{(0)}_\mu}-\tilde{\bm e}^{(0)}\,.
\end{equation}
If we use the transformation rule at the same point~\eqref{Reltetradssamephysicalpoint} we find that at first order
in the transformation the relation between these projectors is given by ~\cite{Tsagas:2007yx}
\begin{equation}\label{RelSStilde}
\tilde S_{\mu \nu}\circ \tilde c{(\tilde{\bm x},q^{\widetilde{(a)}})}=
S_{\mu\nu}({\bm x},q^{(a)})+ 2 \left(e^{(0)}_{(\mu}+n_{(\mu}\right) S_{\nu)
  \alpha}({\bm x},q^{(a)}) V^\alpha\,,\quad \text{with}\quad {\bm V}
\equiv -\Lambda^{(i)}_{\,\,(0)} {\bm e}_{(i)}=-\pa^i T {\bm e}_i\,.
\end{equation}
Note that if we use tetrad coordinates for the argument of the screen projectors, then
\begin{equation}
{\bm S}(q,n^{(i)})={\bm g}+{\bm e}^{(0)} \otimes {\bm e}^{(0)}-n_{(i)}
n_{(j)}{\bm e}^{(i)}\otimes {\bm e}^{(j)}\,,\qquad \tilde{\bm
  S}\circ \tilde c{(q,n^{(i)})}={\bm g}+\tilde{\bm e}^{(0)} \otimes \tilde{\bm
  e}^{(0)}-n_{(i)} n_{(j)}\tilde {\bm e}^{(i)} \otimes  \tilde {\bm e}^{(j)}\,.
\end{equation}
This means that when the coordinates of the projectors are expressed in the tetrad basis associated with the corresponding coordinates system, we obtain
\begin{equation}\label{SijisSij}
S_{(i)(j)}(q,n^{(k)})=\delta_{ij}-n_{(i)}n_{(j)} = \tilde S_{\widetilde{(i)}\widetilde{(j)}}\circ \tilde c{(q,n^{(k)})}\,.
\end{equation}
Thus in any gauge, the expression of the related screen projector in the
tetrad basis is the same by construction, and $S_{(i)(j)}$ depends
actually only on $n^{(k)}$. However it must remain clear that these two tensors are geometrically different since they are
associated with different coordinates systems and indeed their relation is given at first order by~\eqref{RelSStilde}.

The distribution tensor is also not geometrically invariant since for every observer used to define the screen projector, we must consider a different distribution tensor. However all the possible distribution matrices are related through
projections and we find that the distribution tensors defined by the
tetrad $\tilde{\bm e}_{(0)}$ and ${\bm e}_{(0)}$ are related at the same point of the tangent bundle by~\cite{Tsagas:2007yx}
\begin{equation}
f^{\tilde{\bm e}_{(0)}}_{\mu\nu}\circ
 \tilde c{(\tilde {\bm x},q^{\widetilde{(\imath)}})}
 = \tilde S_\mu^\alpha \circ \tilde c{(\tilde {\bm x},q^{\widetilde{(\imath)}})}
 \tilde S_\nu^\beta \circ \tilde c{(\tilde {\bm x},q^{\widetilde{(\imath)}})}
 f^{{\bm e}_{(0)}}_{\alpha \beta}({\bm x},q^{(\imath)})\,.
\end{equation}
This means that the only requirement is to project the distribution
function so that it is projected with respect to the new observer and
the new direction. Combining this transformation rule with~\eqref{RelSStilde}
we obtain at first order the transformation rule as
\begin{equation}
 f^{\tilde{\bm e}_{(0)}}_{\mu\nu}\circ
 \tilde c{(\tilde {\bm x},q^{\widetilde{(\imath)}})}
 = f^{{\bm e}_{(0)}}_{\mu \nu}({\bm x},q^{(\imath)})
 + 2 \left(e^{(0)}_{(\mu}+n_{(\mu}\right)
 f^{{\bm e}_{(0)}}_{\nu)\alpha}({\bm x},q^{(\imath)})V^\alpha\,.
\end{equation}
If we project this expression onto the tetrad components, and noting
$\tilde f_{(i)(j)}\equiv f^{\tilde{\bm e}_{(0)}}_{(i)(j)}$ and $f_{(i)(j)}\equiv f^{{\bm e}_{(0)}}_{(i)(j)}$, we obtain
\begin{equation}\label{Rulefijstrange}
 \tilde f_{\widetilde{(i)}\widetilde{(j)}}\circ
 \tilde c{(\tilde {\bm x},q^{\widetilde{(\imath)}})}
 = f_{(i)(j)}({\bm x},q^{(\imath)})
 - f_{(i)(k)}({\bm x},q^{(\imath)})L_{\,\,\,,j]}^{[k}
 -f_{(k)(j)}({\bm x},q^{(\imath)})L_{\,\,\,,i]}^{[k}
 - 2 n_{((i)} f_{(j))(k)}({\bm x},q^{(\imath)})\pa^k T\,,
\end{equation}
find the transformation rule under a gauge transformation, we need to expand the left
hand side around $({\bm x},q^{(\imath)})$. At first order we get
\begin{align}
\tilde f_{\widetilde{(i)}\widetilde{(j)}}\circ
 \tilde c{(\tilde {\bm x},q^{\widetilde{(\imath)}})}
 &\simeq\left(1+\xi^\mu \frac{\pa}{\pa x^\mu}
 +\delta q^{(i)}\frac{\pa}{\pa q^{(i)}} \right)
 \tilde f_{\widetilde{(i)}\widetilde{(j)}}\circ
\tilde c{({\bm x},q^{{(\imath)}})}+\cdots \nonumber\\
&\simeq \tilde f_{\widetilde{(i)}\widetilde{(j)}}\circ
 \tilde c{({\bm x},q^{{(\imath)}})}
 + \left(\xi^\mu \frac{\pa}{\pa x^\mu}+\delta q^{(i)}\frac{\pa}{\pa q^{(i)}}
 \right) f_{{(i)}{(j)}}+\cdots \,.
\end{align}
Using the expansion~(\ref{Expandderpi}) we then obtain the gauge transformation
rule for the tensor valued distribution function in tetrad
coordinates. Noting $\widetilde{f_{(i)(j)}} \equiv \tilde
f_{\widetilde{(i)}\widetilde{(j)}}\circ \tilde c$ for simplicity, this reads
\begin{equation}\label{Gaugeruletensor}
\widetilde{f_{{(i)}{(j)}}} = f_{(i)(j)}-\left(\xi^\mu \frac{\pa}{\pa x^\mu}+\delta \ln q \parq+\delta n^{(i)}D_{(i)}  \right)f_{(i)(j)}-f_{(i)(k)}
L_{\,\,\,,l]}^{[k}S^{(l)}_{(j)} -f_{(k)(j)} L_{\,\,\,,l]}^{[k} S^{(l)}_{(i)}\,,
\end{equation}
where it is implied that all quantities are evaluated either at ${\bm x}$ or
at $({\bm x},q,n^{(i)})$.

For completeness we report the explicit form of the gauge transformation
 for $\calP_{(i) (j)}$ which is obtained from the above transformation
 rule and the spectral decomposition (\ref{defyP})
 \begin{align}\label{GT:calP}
 \tilde{\calP}_{(i) (j)}
 &= \calP_{(i) (j)}
 - L^{[k}{}_{, l]} \Bigl( S^{(l)}{}_{(i)} \calP_{(k) (j)}
 + S^{(l)}{}_{(j)} \calP_{(k) (i)} \Bigr) - \xi^\mu \pa_\mu \calP_{(i) (j)}
 - \delta n^{(k)} D_{(k)} \calP_{(i) (j)} \,.
 \end{align}
It is also found, as expected, that the spectral distortion $Y_{(i)(j)}$ part is
gauge invariant since it vanishes on the background and at first order.

\subsection{Gauge transformation for Liouville and Collision terms}
\label{AppGTLC}

We deduce from the transformation rule~(\ref{GT:calP}) and
 the spectral decomposition (\ref{spdec:calP_lhs}) and
 (\ref{spdec:calP_rhs}) that the spectral distortion part,
 $\calL^Y_{(i)(j)} $ and $\calC^Y_{(i)(j)}$, must be gauge invariant.
Concerning the temperature part, they should transform as
 (noting $\widetilde{\calL^P_{(i)(j)}}
 \equiv \calL^P_{\tilde{(i)}\tilde{(j)}} \circ \tilde c$ and
 $\widetilde{\calC^P_{(i)(j)}}
 \equiv \calC^P_{\tilde{(i)}\tilde{(j)}} \circ \tilde c$ )
 \begin{align}
 \widetilde{\calL^P_{(i)(j)}}
 &= \calL^P_{(i)(j)} - L^{[k}{}_{, l]} \Bigl( S^{(l)}{}_{(i)} \calL^P_{(k) (j)}
 + S^{(l)}{}_{(j)} \calL^P_{(k) (i)} \Bigr) - \xi^\mu \pa_\mu \calL^P_{(i)(j)}
 - \delta n^{(k)} D_{(k)} \calL^P_{(i) (j)} \,, \\
 \widetilde{\calC^P_{(i)(j)}}
 &= \calC^P_{(i)(j)} - L^{[k}{}_{, l]} \Bigl( S^{(l)}{}_{(i)} \calC^P_{(k) (j)}
 + S^{(l)}{}_{(j)} \calC^P_{(k) (i)} \Bigr) - \xi^\mu \pa_\mu \calC^P_{(i)(j)}
 - \delta n^{(k)} D_{(k)} \calC^P_{(i) (j)} \,.
 \end{align}
Using the transformation rules derived in this paper, we checked that
this is indeed the case when using the detailed form of the Liouville and collision operators.

\section{Extraction of temperature and spectral distortion}
\label{AppExtraction}

The functions $y$ and $\Theta$ can be extracted thanks to
 the integrals of the type
\begin{equation}
 \calM_n[f]\equiv \frac{\int f q^{2+n} \dd q}{(3+n)
 \int \bar I(q) q^{2+n} \dd q} \,,
\end{equation}
just by applying them order by order to $I(q)$, using that
 $\calM_0[\calD_q^2 \bar I]=0$.
We then obtain
\begin{subequations}
\begin{align}
\Theta^{(1)} &= \calM_1 [I^{(1)}] = \calM_0[I^{(1)}] \,,
 \label{ExtractTheta1} \\
 \frac{1}{2}\Theta^{(2)} &= \frac{1}{2} \calM_0[I^{(2)}]-\Theta^{(1)2} \,, \\
 \frac{1}{2}y^{(2)} &= \frac{1}{2} \left( \calM_1[I^{(2)}] - \calM_0 [I^{(2)}]
 \right) - \frac{1}{2}\Theta^{(1)}{}^2 \,.
\end{align}
\end{subequations}

Similarly to what can be done for the intensity part,
 the spectral components of polarisation can be extracted thanks to
\begin{subequations}
\begin{align}
 \calP^{(1)}_{\mu\nu}
 &= \calM_1 [{\bf P}_{\mu\nu}^{(1)}] = \calM_0 [P_{\mu\nu}^{(1)}] \,, \\
 \frac{1}{2} \calP^{(2)}_{\mu\nu}
 &= \frac{1}{2} \calM_0 [P_{\mu\nu}^{(2)}]
 - 3 \Theta^{(1)} \calP^{(1)}_{\mu\nu} \,,\\
 \frac{1}{2}Y^{(2)}_{\mu\nu}
 &= \frac{1}{2} \Bigl( \calM_1[P_{\mu\nu}^{(2)}]-\calM_0[P_{\mu\nu}^{(2)}] \Bigr)
 - \Theta \calP^{(1)}_{\mu\nu} \,.
\end{align}
\end{subequations}



\end{document}